\documentclass{aa}
\usepackage[varg]{txfonts}
\usepackage{graphicx}
\usepackage{natbib}
\usepackage{subfigure}
\usepackage{rotating}
\usepackage{lscape}
\bibpunct{(}{)}{;}{a}{}{,}

\begin{document}

\title{The Fornax Deep Survey (FDS) with VST \\ 
XII: Low surface brightness dwarf galaxies in the Fornax cluster }

\author{Aku Venhola\inst{1}
        \and Reynier F. Peletier \inst{2} 
        \and Heikki Salo\inst{1}
        \and Eija Laurikainen\inst{1} 
        \and Joachim Janz\inst{3}$^{,1}$
        \and Caroline Haigh\inst{4}
        \and Michael H. F. Wilkinson\inst{4}
        \and Enrichetta Iodice\inst{5}
        \and Michael Hilker\inst{6}
        \and Steffen Mieske\inst{7}
        \and Michele Cantiello\inst{8}
        \and Marilena Spavone\inst{5}
}

\offprints{A. Venhola, \email{aku.venhola@oulu.fi}}

\institute{Space physics and astronomy research unit, University of Oulu, Finland 
  \and Kapteyn Institute, University of Groningen, Groningen, the Netherlands
  \and Finnish Centre of Astronomy with ESO (FINCA)
, University of Turku, Väisäläntie 20, FI-21500 Piikkiö, Finland  
  \and Bernoulli Institute of Mathematics, Computer Science and Artificial Intelligence, University of Groningen, Groningen, the Netherlands
  \and INAF - Astronomical Observatory of Capodimonte, Salita Moiariello 16, I-80131, Naples, Italy
  \and European Southern Observatory, Karl-Schwarzschild-Strasse 2, D-85748 Garching bei Muenchen, Germany
  \and European Southern Observatory, Alonso de Cordova 3107, Vitacura, Santiago, Chile
  \and INAF-Astronomical Abruzzo Observatory, Via Maggini, 64100, Teramo, Italy}

\date{Received \today / Accepted ---}

\abstract { Low surface brightness (LSB) dwarf galaxies in galaxy clusters are an interesting group of objects as their contribution to the galaxy luminosity function and their evolutionary paths are not yet clear. Increasing the completeness of our galaxy catalogs is crucial for understanding these galaxies, which have effective surface brightnesses below 23 mag arcsec$^{-2}$ (in optical). Progress is continuously being made via the performance of deep observations, but detection depth and the quantification of the completeness can also be improved via the application of novel approaches in object detection. For example, the Fornax Deep Survey (FDS) has revealed many faint galaxies that can be visually detected from the images down to a surface brightness level of 27 mag arcsec$^{-2}$, whereas traditional detection methods, such as using Source Extractor (SE), fail to find them.
} 
{ In this work we use a max-tree based object detection algorithm (Max-Tree Objects, MTO) on the FDS data in order to detect previously undetected LSB galaxies. After extending the existing Fornax dwarf galaxy catalogs with this sample, our goal is to understand the evolution of LSB dwarfs in the cluster. We also study the contribution of the newly detected galaxies to the faint end of the luminosity function. 
}
{
 We test the detection completeness and parameter extraction accuracy of MTO using simulated and real images. We then apply MTO to the FDS images to identify LSB candidates. The identified objects are fitted with 2D S\'ersic models using GALFIT and classified as imaging artifacts, likely cluster members, or background galaxies based on their morphological appearance, colors, and structure.
}
{  With MTO, we are able to increase the completeness of our earlier FDS dwarf catalog (FDSDC) 0.5-1 mag deeper in terms of total magnitude and surface brightness. Due to the increased accuracy in measuring sizes of the detected objects, we also add many small galaxies to the catalog that were previously excluded as their outer parts had been missed in detection.  We detect 265 new LSB dwarf galaxies in the Fornax cluster, which increases the total number of known dwarfs in Fornax to 821. Using the whole cluster dwarf galaxy population, we show that the luminosity function has a faint-end slope of $\alpha$=-1.38$\pm$0.02. We compare the obtained luminosity function with different environments studied earlier using deep data but do not find any significant differences. On the other hand, the Fornax-like simulated clusters in the IllustrisTNG cosmological simulation have shallower slopes than found in the observational data. We also find several trends in the galaxy colors, structure, and morphology that support the idea that the number of LSB galaxies is higher in the cluster center due to tidal forces and the age dimming of the stellar populations. The same result also holds for the subgroup of large LSB galaxies, so-called ultra-diffuse galaxies.
} {}

\maketitle   

\section{Introduction}

In the last decade, the deployment of wide field imaging facilities with large collecting areas has improved our ability to study low surface brightness (LSB) galaxies\footnote{We define an LSB galaxy as a galaxy with an r'-band mean effective surface brightness $\bar{\mu}_{e,r'}$ $\gtrsim$ 23 mag arcsec$^{-2}$.}. Several deep surveys, such as the Next Generation Virgo Survey (NGVS; \citealp{Ferrarese2012}), the Kilo Degree Survey (KiDS; \citealp{deJong2015}), the VST Early-type GAlaxy Survey (VEGAS; \citealp{Capaccioli2015}), and the Fornax Deep Survey (FDS; \citealp{Iodice2016}, \citealp{Venhola2018}, hereafter V18), have gathered large amounts of data that reveal previously unseen LSB galaxies (\citealp{Munoz2015}, \citealp{Koda2015}, \citealp{VanDerBurg2016}, \citealp{Chamba2020}, \citealp{Prole2019}). These galaxies, having an extremely diffuse structure and existing in dense environments, allow us to test our galaxy formation theories in a so far unexplored and unique parameter space. Ultimately, the number and properties of these galaxies will allow us to probe their formation mechanisms and can be a benchmark for our cosmological models.

\indent So far, we have not reached any lower limit in surface brightness (SB) where we stop finding new galaxies (see, e.g., \citealp{Fattahi2020}). In addition to not knowing how many galaxies we fail to detect, spectroscopic instruments cannot reach similar depths as effectively as the imaging instruments. This limits our understanding of LSB galaxies -- the confirmation of galaxy cluster membership by radial velocities is made difficult, and thus the contribution of the LSB galaxies to the galaxy luminosity function (LF), and their properties in general, is not well known. As there exists a luminosity-SB relation for galaxies (\citealp{Binggeli1984}, \citealp{Kormendy2012}), most of the galaxies that have low SBs are low-mass dwarfs, but massive LSB galaxies also exist (\citealp{Romanishin1983}, \citealp{Bothun1985}, \citealp{Impey1989}, \citealp{Impey1997}).

\indent  Massive LSB galaxies are typically blue, slowly star-forming spiral galaxies \citep{McGaugh1994}. In the case of dwarfs, there are both blue star-forming and red quiescent LSB dwarf galaxies (\citealp{Leisman2017}, \citealp{Roman2017}). Low surface brightness dwarfs follow the morphology-density relation \citep{Dressler1980}, such that the red LSB dwarfs are mostly in dense environments and the blue LSB dwarfs in sparse ones (\citealp{Leisman2017}, \citealp{Roman2017}). In this study we concentrate on the analysis of the low-mass LSB galaxies in the Fornax cluster. Thus, most of the galaxies are expected to be quiescent and gas poor.

\indent A subgroup of LSB dwarf galaxies called ultra-diffuse galaxies (UDGs; \citealp{Sandage1984}, \citealp{VanDokkum2015}), which are defined by having large effective radii (R$_e$ $>$ 1.5 kpc), has been studied in several environments in order to understand what causes their large sizes and low SBs (\citealp{DiCintio2016}, \citealp{Zaritsky2019}, \citealp{Merrit2016}, \citealp{Mancera2018}, \citealp{VanDerBurg2016}, \citealp{Mihos2015}, \citealp{Venhola2017}, \citealp{Wittmann2016}, \citealp{Iodice2020}, \citealp{Leisman2017}). These studies have found little evidence of UDGs constituting a distinct population of dwarf galaxies (\citealp{Conselice2018}, \citealp{Koda2015}, \citealp{Chilingarian2019}, \citealp{Mancera2019}, \citealp{Lim2020}); they instead point to them being the diffuse end in the continuum of dwarf galaxy properties.

\indent Efforts to try to explain the origin of UDGs have led to progress in understanding different external and internal mechanisms that make some dwarf galaxies of a given mass have lower SBs than others. Ultra-diffuse galaxies and other LSB dwarfs tend to form naturally in modern simulations: The large angular momentum of dark matter (DM) halos and gas leads to large galaxies with low SBs (\citealp{Amorisco2016}, \citealp{Tremmel2020}). The dependence between the rotation of the DM halos and the density of the environment might lead to an increased fraction of LSB dwarf galaxies in high density environments \citep{Tremmel2020}.  However, star formation suppression due to feedback can also cause the low SB of the low-mass galaxies (\citealp{DiCintio2016}, \citealp{DiCintio2019}, \citealp{Jiang2019}), and not all simulations find a dependence between the SB and the rotation of the DM halo or the stellar body of low-mass galaxies (\citealp{Jiang2019}, \citealp{DiCintio2019}). Thus, it seems that there are different mechanisms that can lead to the formation of LSB galaxies independent of the density of their formation environment.

\indent According to simulations, some LSB dwarf galaxies also form during and after the infall of galaxies into a cluster. Quenching of the star formation by ram-pressure stripping (RPS) during the infall of dwarf galaxies leads to passively aging stellar populations. After the brightest and most massive stars end their nuclear burning phase, the SB of the galaxies fades over time if no new stars are born. This leads to some dwarf galaxies with high SB during their infall to a denser environment fading into LSB dwarfs and UDGs (\citealp{Tremmel2020}, \citealp{Jiang2019}). This process also leads to the SB-age relation predicted by simulations \citep{Tremmel2020}. As the galaxies are dragged toward the center of the cluster over time due to tidal friction, galaxies that have fallen into a cluster early are now likelier to appear as LSB galaxies near the center of the cluster relative to the recently in-fallen population (\citealp{Tremmel2020}, \citealp{Jiang2019}). Hence, the fraction of LSB dwarf galaxies with respect to non-LSB dwarfs should increase toward the cluster center \citep{Tremmel2020}. In addition to the decrease in SB due to the aging of the stellar populations, the surface density (SD) of galaxies that travel near the cluster center during their pericenter can also decrease by as much as 30\% due to the tidal field of the cluster (\citealp{Tremmel2020}, \citealp{Sales2019}). This decrease in the SD happens when DM is stripped from the halo of the galaxies and, as a result, the stellar body of the galaxy expands due to the changes in the gravitational potential.

\indent From an observational point of view, a lot of work has been done in order to identify UDGs and other LSB dwarfs. Their distribution and abundance has been analyzed within and outside of groups and clusters. The number of UDGs is found to be proportional to the mass of the cluster (or group), and they appear to follow a cored spatial distribution in the cluster similar to that of other dwarf galaxies (\citealp{VanDerBurg2016}, \citealp{Mancera2018}, \citealp{Wittmann2016}). The concentration of the UDGs' light distribution and their projected shapes are very similar to those of dwarf galaxies of the same mass but higher SB (\citealp{Mancera2019}). They also have a bimodality in their color distribution similar to other dwarf galaxies (\citealp{Roman2017}). From spectroscopic studies of small samples of UDGs, we know that their stellar populations and kinematics are similar to those of other dwarfs (\citealp{Chilingarian2019}, \citealp{RuizLara2018}).

\indent Regardless of the effort in identifying UDGs and other LSB dwarfs and comparing their properties with other galaxies in other environments, less work has been done to set the LSB dwarfs in the context of a complete dwarf galaxy population in any given environment. This is clearly necessary in order to understand how the LSB dwarfs are different from other galaxies. With a complete dwarf galaxy population, it is straightforward to test whether we find the observable relations found in the simulations, for example, an age-SB relation, a correlation between the cluster-centric distances and sizes of the galaxies, or differences between the distribution of UDGs and other dwarfs.

\indent Recently, Lim et al. (2020) studied the UDGs in the Virgo cluster within a complete galaxy sample and showed that they are a normal diffuse end tail of the cluster dwarf galaxy population. Contrary to what was thought based on previous studies (e.g., \citealp{Mancera2018}), they also found that the UDGs are more centrally clustered in the cluster than other dwarfs, in agreement with simulation predictions  \citep{Sales2019}. Their study clearly demonstrated how important it is to place UDGs in the context of a complete galaxy population.

\indent In order to generate a useful galaxy catalog for the abovementioned purpose, we need to identify and quantify galaxies in a robust, homogeneous manner. Many galaxy catalogs compiled from the data of new surveys are produced by identifying the galaxies either visually or automatically using Source Extractor (SE; \citealp{Bertin1996}). The problems related to those methods are well known and have been pointed out by, for example, \citet{Akhlaghi2015} and \citet{Venhola2017}. At present, many better methods have become available, such as Max-Tree Objects (MTO; \citealp{Teeninga2016}),  NoiseChisel \citep{Akhlaghi2015}, and ProFound \citep{Robotham2018}. A quantitative comparison between the novel detection algorithms by \citet{Haigh2021} showed that, when optimized for detection completeness and accuracy of the measured parameters, MTO slightly outperforms NoiseChisel and others. These improved algorithms have not yet been widely used; for example, the statistical studies of UDGs (\citealp{Yagi2016}, \citealp{VanDerBurg2016}, \citealp{Mancera2018}) are all based on catalogs obtained with SE. However, some new studies have already started using, for example, MTO (\citealp{Prole2019}, \citealp{Chamba2020}, \citealp{Muller2021}). With the new detection algorithms, we can reduce the systematic biases in the LSB samples and, moreover, obtain more complete catalogs from the currently available data sets.

In the ground-based observations of LSB dwarfs, seeing often sets a limit for the minimum size of the objects that we can separate from background galaxies. Hence, studying a nearby galaxy cluster allows us to resolve small galaxies best. Being the second closest cluster to us and being covered with deep, publicly available survey data, the Fornax cluster is an excellent environment for studying the faintest dwarfs. The basic properties of the Fornax cluster are listed in Table \ref{tab:Fornax}. 

\begin{table}[]
    \centering
    \begin{tabular}{c c c}
     & Value & Source \\  \hline
    Mass (M$_{\odot}$) & 7$\pm$2 $\times$ 10$^{13}$ & 1 \\
    Virial radius (Mpc) & 0.7 & 1\\
    Virial Radius (deg) & 2.0 & 1 \\
    Velocity dispersion (kms$^{-1}$) & 318 & 2\\
    Distance modulus (mag) & 31.51 & 3
    \end{tabular}
    \caption{Properties of the Fornax cluster. Explanations for the sources: 1) \citet{Drinkwater2001}, 2) \citet{Maddox2019}, 3) \citet{Blakeslee2009}. }
    \label{tab:Fornax}
\end{table}

\indent In this study we apply MTO to detect LSB dwarf galaxies in the Fornax cluster. A few studies have mapped the LSB galaxy population in the Fornax cluster before: \citet{Bothun1991}, \citet{Hilker1999}, \citet{Hilker2003} \citet{Mieske2007}, \citet{Munoz2015}, \citet{Venhola2017}, \citet{Eigenthaler2018}, \citet{Ordenes-Briceno2018}, and V18. All of these studies identify objects either visually or by applying SE and are mostly concentrated on the central area of the cluster. In this work we aim to extend the FDS dwarf galaxy catalogs in order to obtain a homogeneous and complete sample of dwarf galaxies. \citet{Venhola2017} visually studied the LSB galaxy population in the 4 deg$^2$ area around the center of the cluster from the FDS data. The faintest objects in that catalog reach $\bar{\mu}_{e,r'}$ $\approx$ 28 mag arcsec$^{-2}$ and have sizes of up to R$_e$ $\approx$ 10 kpc, making it the most complete compilation in that area of the Fornax cluster. V18 published a spatially complete catalog of dwarfs for the whole  26 deg$^2$ area of the FDS using SE for the detection, but that catalog misses many LSB galaxies. Throughout the paper, we refer to these works as the core LSB catalog \citep{Venhola2017} and the Fornax Deep Survey Dwarf Catalog (FDSDC; V18). In this work we aim to increase the LSB completeness of the FDSDC to a depth similar to that of the core LSB catalog.

\indent In this study we first describe the FDS and a simulated data set (Sect. 2) that we use for the quality assessment of our object detection (Sect. 3). We apply MTO to detect LSB galaxies in the full FDS data set and fit light profiles of the detected galaxies using GALFIT \citep{Peng2010} (Sect. 4). The resulting LSB candidates are separated into cluster members and background objects (Sect. 5), and the distribution and properties of the cluster LSB galaxies are analyzed (Sect. 6). We then discuss how the newly identified LSB dwarf population relates to the previously known dwarf populations in the Fornax cluster and other environments and simulations in Sect. 7, and we summarize our results  in Sect. 8.

\indent Throughout the paper we use the distance of 20.0 Mpc for the Fornax cluster, which corresponds to the distance modulus of 31.51 mag \citep{Blakeslee2009}. At this distance, 1 deg corresponds to 0.349 Mpc and 1 arcsec corresponds to 100 pc.

\section{Data and existing catalogs}

We aim to detect as many as possible bona fide LSB dwarf galaxies from the FDS data set. Moreover, in order to test the completeness and biases of our detection algorithm, we require a data set with a known ground truth and similar image quality to the FDS. For that purpose we generated artificial images that resemble the FDS images. These data sets are described in the following subsections.

\subsection{The Fornax Deep Survey data}

The ESO VLT Survey Telescope \citep[VST;][]{Schipani2012} observations of the FDS cover a 21 deg$^2$ area centered on the Fornax main cluster in the u', g', r' and i' bands, and an additional 5 deg$^2$ area centered on the infalling Fornax A subgroup, observed in the g', r', and i' bands. The data are collected with the OmegaCAM instrument, which is a 1 deg$^2$ field-of-view 32-CCD camera with a 0.21 arcsec pixel size\footnote{The pixel size was re-binned to 0.2 arcsec$^2$ during the mosaicing of the data.}. The reduction and calibration steps of the data are described by V18.

\indent The photometric uncertainties of the data were estimated by V18, and they are 0.04, 0.03, 0.03, and 0.04 mag in u', g', r' and i', respectively. Fields are covered with a nearly homogeneous depth with the 1$\sigma$ limiting SB over 1 pixel (0.2 arcsec $\times$ 0.2 arcsec) of 26.6, 26.7, 26.1 and 25.5 mag arcsec$^{-2}$ in u', g', r' and i', respectively. When averaged over 1 arcsec$^{2}$ those limits correspond to limiting SBs of 28.3, 28.4, 27.8, 27.2 mag arcsec$^{-2}$ in u’, g’, r’, and i’, respectively.

\indent The FDS provides two photometric galaxy catalogs of dwarfs and separate measurements for the giant galaxies ( \citealp{Iodice2019}, \citealp{Spavone2020}, \citealp{Raj2019}, and \citealp{Raj2020}). The FDSDC, which contains all the dwarf galaxies with semimajor axis $a$ $>$ 2 arcsec in the whole FDS, was generated by V18.  The 50 \% completeness limit of this catalog is at the r'-band mean effective SB of $\bar{\mu}_{e,r'}$ $\approx$ 26 mag arcsec$^{-2}$. \citet{Venhola2017} presented an LSB galaxy catalog that was made by visually identifying LSB galaxies from the 4 deg$^2$ in the center of the cluster. For the next steps we assume that this core LSB catalog is complete within the depth limits of the data, as it was made by inspecting the images manually and selecting all LSB objects. In this work we utilize the core LSB catalog and FDSDC for testing the detection completeness obtained when using MTO and for the base catalogs in the analysis.

\subsection{Simulated images}

For generating the mock images, we used the background noise and point spread function (PSF) model of the r'-band FDS field 11 described by V18. The artificial observations were produced by first generating a blank image with a size of 10,000 $\times$ 10,000 pixels (corresponding to approximately a quarter of a single FDS field). 

\indent We then injected point sources representing stars to the image, using an apparent luminosity distribution drawn from an exponential distribution N$\propto$ $\exp($m$_{r'}$/3$)$ between object magnitudes of 12 mag $<$ m$_{r'}$ $<$ 27 mag. This distribution and the number density of the stars was selected to be similar to that found from the FDS images based on a SE detection of stars in the Field 11. We then embedded S\'ersic models at random positions to the image as mock galaxies, which we consider to be  representative for both the cluster and background galaxies. The number of background and cluster galaxies were 4000 and 50, respectively, per image. The structural parameter ranges of the mock galaxies were matched to the values found for the cluster and background galaxies\footnote{For the cluster galaxies, we used a uniform distribution for SB 21 mag arcsec$^{-2}$ < $\mu_{e,r'}$ < 31 mag arcsec$^{-2}$, S\'ersic index 0.5 < $n$ < 2, and axis ratio 0.3 < $b/a$ < 1. For effective radii, we used a range of 1 arcsec < R$_e$ < 300 arcsec, so the sample also includes some very large LSB galaxies. For the background galaxies we used 21 mag arcsec$^{-2}$ < $\mu_{e,r'}$ < 31 mag arcsec$^{-2}$, 2 < $n$ < 4, and axis ratio 0.3 < $b/a$ < 1 and  0.5 arcsec < R$_e$ < 4 arcsec.} in the catalogs of V18, but the SB distribution was extended down to $\bar{\mu}_{e,r'}$ = 31 mag arcsec$^{-2}$.

\indent  After embedding all the objects in units of electron counts, the image was convolved with the OmegaCAM r'-band PSF (V18), and the Poisson noise was applied. Finally, Gaussian background noise was added to the pixels, and the pixel values were divided by the gain factor. We generated 110 such mock images, which we used to assess the detection completeness of MTO and the accuracy of the photometric measurements (see Sects. 3.2 and 4.1).

\indent We acknowledge that these simple mock images cannot reproduce all the complexity of the real data (e.g., different galaxy morphologies, globular clusters in Fornax galaxies, reduction artifacts, PSF variations, reflections, satellite tracks etc.), which might lead to some biases in the analysis of the completeness limits. To ensure that we nevertheless obtain good estimates for the completeness limits we also test the detection with real images (using the core LSB catalog). However, the main benefit of using the simulated data is that we can be certain about the ground truth of the images, and thus we are able to objectively quantify the accuracy of the detections and the photometry.

\section{LSB object identification}

We use MTO for identifying LSB objects. MTO is an object detection program that utilizes maximum trees (\citealp{Salembier1998}, \citealp{Souza2016}) for identifying structures in an image and keeps track of the hierarchical structure of the identified objects. This approach allows the program to deal with nested objects (i.e., objects that appear on top of each other), which is especially important in the cluster environment where the object density is high. 

\indent \citet{Haigh2021} have shown that MTO outperforms other common source detection programs in completeness and purity. In their paper, Haigh et al. used a set of simulated and real images with known ground truths to test NoiseChisel, Profound, SE, and MTO. They optimized the programs using a combined completeness and area score, and found that MTO obtains highest score for both measures. They also found that MTO does not require a tuning of parameters when changing from a data set to another, whereas for other algorithms the optimal set of parameters changed between the data sets. A caveat of MTO is that it is much slower than SE: It takes about 20 mins to analyze one FDS field (21 kpix $\times$ 21 kpix) with MTO using a typical desktop computer, whereas with SE it takes less than half a minute. However, as the entirety of the FDS data is only tens of gigapixels, the slower processing time is not a problem. As Haigh et al. have done these comparisons in detail (see also \citealp{ThesisVenhola} Ph.D. thesis), we do not perform further comparisons between the other different detection algorithms, instead focusing on testing the performance of MTO using the simulated and original FDS data and comparing it with the results obtained when using SE\footnote{SE is interesting here as it was used for object detection in the FDSDC.}.

\subsection{Detection of LSB galaxies using MTO}

The algorithm of MTO has been described in detail by \citet{Teeninga2016}. However, for completeness we give a short overview of the detection process here.

\indent  MTO assumes that an input image, with a gain of $g$, consists of three different components (all in fluxes): Gaussian background noise with variance $g^{-1}B$, a Poisson distributed signal $g^{-1}O$, and other noise sources $R$, such as readout noise. The background is assumed to be flat over the whole image, thus being well approximated by the mean of the pixel values of the pixels devoid of object signal. The five main steps of the algorithm are as follows.

The first is background estimation: The image is first divided into 64$\times$64 pixel (12.8 $\times$ 12.8 arcesc$^2$) tiles. In each tile statistical tests are made to test the Gaussianity and flatness of the tiles. The Gaussianity test is done using the D'Agostino-Pearson K$^2$-statistic. Then the tiles are tested for absence of gradients (flatness) by comparing the means within the different tile quadrants with each other using a t-test. Depending on whether any tiles pass the background tests, the common tile size is increased or decreased by a factor of two. The background size is then selected to be the largest scale in which tiles pass those tests. The tiles passing these two tests are then selected to represent the background, and the mean value of those tiles is subtracted from the image. Background variance $g^{-1}B$ is also measured from the selected tiles. The background becomes the first node of the tree.

The second is the identification of significant branches: The image is thresholded with increasing threshold levels, and at each level the maximum connected structures are identified. The significance of each structure is tested by calculating its power\footnote{The power of a branch consisting of the pixel set $P$ is defined as $Power(P)=\Sigma_{x \in P}\left( f(x)-A\right)^2$, where $f(x)$ returns the flux in pixel $x$ and $A$ is the intensity level of the parent node.}, and it is considered as a significant branch if its power exceeds the limit $\alpha$. We used the default value for $\alpha$ that rejects the detections that are less significant than 5$\sigma$. The hierarchical structure of the branches is then stored into a tree.

The third is the labeling of the branches: The branches in the tree are labeled so that the branches that belong to the same object obtain the same label. In the case of nodes (i.e., several branches sharing the same parent), the branch with the most power obtains the same label as its parent, and all the other branches obtain new labels. 

The fourth is the generation of the segmentation map: A segmentation map is produced by projecting the tree into a plane with same dimensions as the input image. The labels for the segmentation map are taken from the highest level branch that is present in the given pixel.

The fifth is parameter extraction: Structural parameters for each object are measured from the input image. These parameters are calculated using the pixels that have the corresponding label in the segmentation image. For nested objects, the intensity level of the parent node is subtracted from the object's pixel values before calculating the parameters. This works as a local background estimation and thus prevents the measured SB of the nested objects from being biased by their parent objects. Parameters that we use in this work are the object center coordinates that are calculated from the flux weighted mean positions of the pixels. Total magnitudes are calculated as a sum of the flux in those pixels. For the measure of size, we use the length of semiminor and semimajor axes, which are calculated from the pixel value distribution moments (see Appendix A.1).

For a given input image, MTO outputs a segmentation map and an object list that includes the locations of the detected objects and their magnitude, size, and median SB.

\subsection{Detection completeness}

In this section we assess the detection completeness and parameter extraction accuracy of MTO using artificial and real images (described in Sects. 2.1 and 2.2). In detail, we test the detection completeness of MTO and the accuracy of the output parameters of the detected objects. The latter test is crucial despite the fact that we will later perform a more elaborate photometry for the objects since the MTO magnitudes and effective radii of the detected objects will be used when the objects of interest are selected among the large number of detections.

\subsubsection{Completeness with mock galaxies}

We ran MTO for the 110 simulated fields (see Sect. 2.2) and compared the detections with the known locations and parameters of the simulated galaxies (i.e., the ground truth). In order to have a comparison for the quality assessment, we also ran the detections using SE. For SE, we optimized the detection parameters by running the SE-detection for simulated images using different parameters and chose the set of parameters that produced the highest $F-score$. Like \citet{Haigh2021}, we decided to use $F-score$ for the optimization as it takes into account both $completeness$ (the proportion of objects that are detected) and $precision$ (the proportion of detections that can be matched to real objects). $F-score$ is defined as follows:
\begin{equation}
    F-score\, = \, 2 \times \frac{precision \times completeness}{precision + completeness}.
\end{equation}

The parameter values we obtained using the optimization are similar to those obtained by \citet{Haigh2021}, although we used larger background size than they did. Details of the set of parameters tested and the used parameters are listed in the Appendix A.2.

\indent When counting the number of detections, we defined a mock galaxy to be detected if there is an object in the output catalog with semimajor axis $a$ $>$ 2 arcsec and $a$ $>$ 0.1$\times$R$_{e,in}$ located within half R$_e$ from the center of the embedded mock galaxy\footnote{ We adopted the $a$ $>$ 2 arcsec selection limit in this test and later in the preparation of the catalog as the typical PSF full width half maximum of the FDS is around 1 arcsec. This selection criterion thus excludes stars and unresolved background galaxies. The purpose of the $a$ $>$ 0.1$\times$R$_{e,in}$ selection limit is to reduce false positive detections from small LSB objects residing on top of extended LSB objects; this rule only concerns a minor portion of objects.}. In the case of nested galaxies or overlapping stars, this rule leads to some false positive detections, but by inspecting the detections we found that this effect is minor (few percent). 

\begin{figure*}
        \includegraphics[width=17cm]{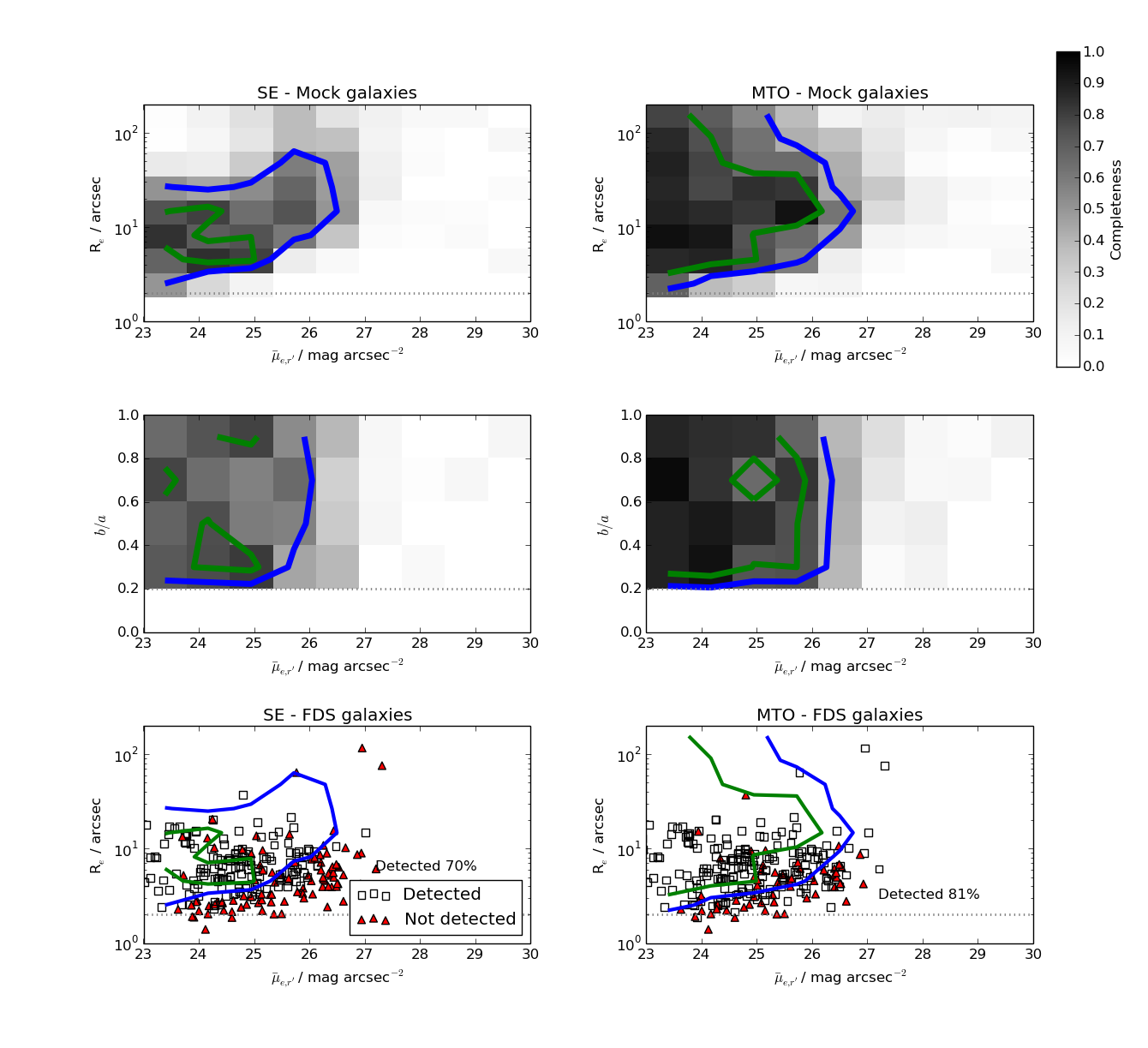}
    \caption{Detection completeness of SE (left panels) and MTO (right panels) as a function of the input r'-band mean effective SB ($\bar{\mu}_{e,r}$), effective radius (R$_e$), and axis ratio ($b/a$).  The upper and middle panels show the results for the artificial images and the lower panels for the known core LSB catalog galaxies (\citealp{Venhola2017}). The gray scales in the upper and middle panels correspond to the completeness indicated by the color bar. The green and blue contours show the 75 \% and 50 \% completeness limits, respectively. Those same lines are also shown in the lower panels. In the lower panels, squares show the detected galaxies and red triangles show the non-detections.}
        \label{fig:completeness_sex_mto}
\end{figure*}

\indent Completeness of the SE and MTO detections are shown as a function of the input r'-band mean effective SB ($\bar{\mu}_{e,r}$) and effective radius (R$_e$) in the upper panels of Fig. \ref{fig:completeness_sex_mto}. Compared to SE, MTO extends the detection completeness both toward larger and smaller R$_e$ and toward fainter SB. The improvements in the detection of galaxies with lower SB and larger size are expected due to the improved background estimation of MTO with respect to SE and the continuous thresholding (\citealp{Teeninga2016}). The improvement in the completeness of objects with small R$_e$ is caused by the selection criteria of objects: more accurate detection of object outskirts leads to more accurate size estimations for the objects. If an object with low SB and small R$_e$ is detected only in its inner parts, the resulting size is underestimated and the object is then excluded by the selection limits due to its small size.

\indent We tested the detection completeness also against the projected shape of the galaxy (see Fig. \ref{fig:completeness_sex_mto}). We find no preferable selection of elongated nor round galaxies. In the Appendix A.3, we show the detection completeness as a function of all the input structural parameters (Fig. \ref{fig:full_Det_an}). We found no significant biases except that the very large and very small R$_e$ galaxies are poorly detected for $\bar{\mu}_{e,r'}$ > 25.5 mag arcsec$^{-2}$.


\begin{figure*}
        \includegraphics[width=17cm]{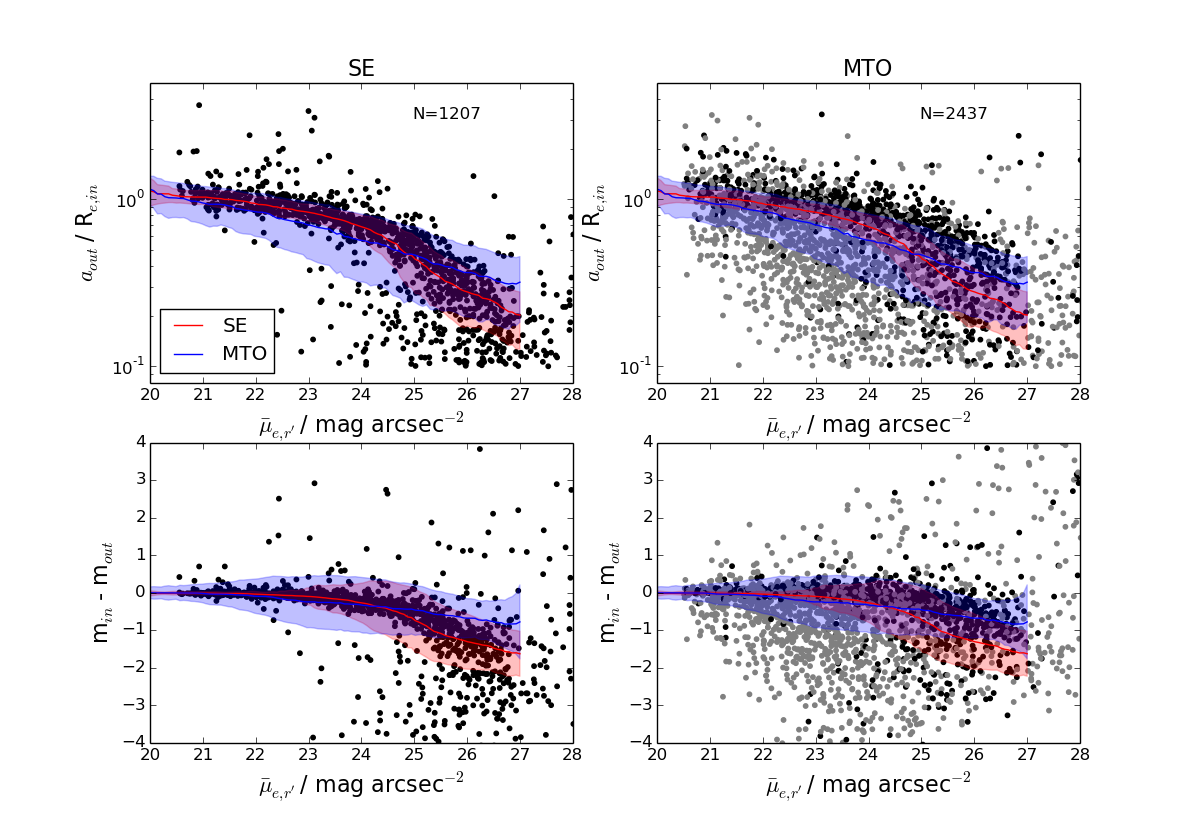}
    \caption{Upper panels: Ratio between the measured semimajor axis length and the input effective radii ($a_{out} / $R$_{e,in}$) for the mock galaxies as a function of their mean effective SB in the r' band ($\bar{\mu}_{e,r'}$). Lower panels: Difference between the input magnitudes (m$_{in}$) and the measured magnitudes (m$_{out}$). The left and right panels show the comparisons when using SE and MTO, respectively. The blue (MTO) and red (SE) lines correspond to the running means of the measurements with a filter size of $\Delta \bar{\mu}_{e,r'}$ = 1 mag arcsec$^{-2}$. Shaded areas around the running means show the median absolute deviation from the means with the corresponding colors. The gray and black symbols in the right panels correspond to the objects that are only detected using MTO and to the objects detected using both methods, respectively. }
        \label{figparams_in_out}
\end{figure*}

\begin{figure*}
        \includegraphics[width=13cm]{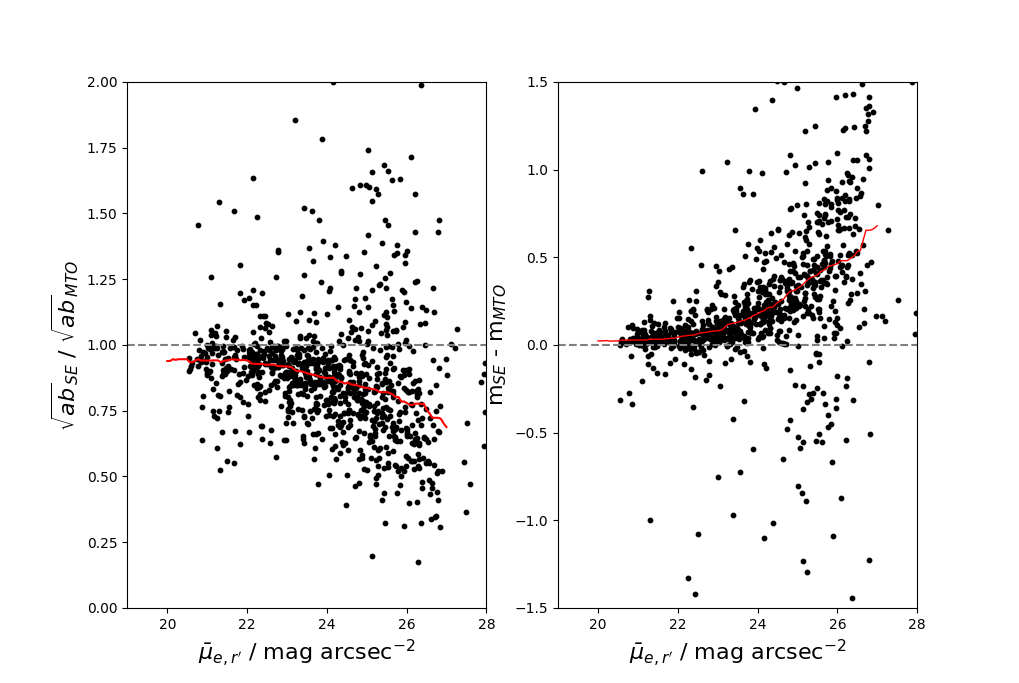}
    \caption{Comparison of the parameters extracted by SE and MTO for the mock galaxies detected by both methods. The left panel shows the ratio between the circularized radii of the objects obtained by SE ($\sqrt{ab}_{SEx}$) and by MTO ($\sqrt{ab}_{MTO}$) as a function of the input r'-band mean effective SB ($\bar{\mu}_{e,r'}$). The right panel shows the difference between the apparent magnitudes obtained with SE (m$_{SEx}$) and MTO (m$_{MTO}$). The red lines show the running medians of the measurements with a filter size of $\Delta \bar{\mu}_{e,r'}$ = 1 mag arcsec$^{-2}$. }
        \label{fig:params_mto_sex}
\end{figure*}

\indent As we know the input effective radii and magnitudes of the mock galaxies, we can compare those with the parameters extracted by MTO and SE. With MTO and SE we did not measure the effective radii but rather the semimajor axes based on second-order moments of the pixel value distribution of the objects (see Appendix A.1). This means that we need to compare those two different measures of size in order to understand how well the extents of the objects are measured. We note that for typical dwarf galaxies with exponential profiles, R$_e$ $\approx$ $a$ (see Fig. A.1).  In Fig. \ref{figparams_in_out}, we show the comparison of the input versus output values of both methods as a function of the input SB of the objects. The trends for the two methods are qualitatively similar showing a decrease in the extracted object sizes and apparent magnitudes toward lower SBs. Both of these methods underestimate the sizes and measure too faint apparent magnitudes for the lowest SB objects, but SE underestimates the values more severely. In the case of MTO, the systematic underestimation of luminosity goes from 0.1 mag to 1 mag in the $\bar{\mu}_{e,r'}$ range from 24 to 27 mag arcsec$^{-2}$.  It is also important to take into account the fact that MTO detects approximately twice the number of objects the SE detects, which means that its trends get affected by objects that SE fails to detect. To take this into account, in Fig. \ref{fig:params_mto_sex} we show the comparison of the parameters obtained by SE and MTO for the objects that both programs have detected. Clear trends are visible both in object sizes and apparent magnitudes, so SE measures fainter magnitudes and smaller sizes for the objects than MTO. Those differences become especially significant for the objects fainter than $\bar{\mu}_{e,r'}$ $>$ 24 mag arcsec$^{-2}$ since there is $\approx$ 30\% systematic difference in the sizes and 0.5 mag difference in the apparent magnitudes.

\indent In summary, we found that MTO obtains approximately one magnitude deeper completeness limits than SE. Additionally, it measures the size and luminosity parameters of the galaxies more accurately than SE. Despite the fact that MTO improves the accuracy of extracted magnitudes with respect to SE, it still measures magnitudes that are 1-2 mag too faint for many objects (lower-right panel of Fig. \ref{figparams_in_out}), and thus more accurate photometric measurements are needed for the selected objects. At least two factors contribute to the bias in the MTO photometry: The outer parts of the objects that are nested with brighter objects will be associated with the brighter ones as no models are assumed for the objects. Additionally, the outer parts of the faintest objects are lost in the noise.

\subsubsection{Completeness with the FDS data}

In order to test the detection completeness also with real data and confirm the completeness limits that we obtained, we use the core LSB catalog. This catalog is assumed to reach the depth limits of the FDS data, as it contains all the LSB galaxies that can be seen in the images by visual inspection. 

\indent In order to perform a quantitative test, we run MTO and SE for the 4 deg$^2$ area in the center of the Fornax cluster that is covered both by the core LSB catalog and the FDSDC. We then compare the object lists produced by each of the methods with those catalogs. We adopted the same rules for counting the detections as we did with the artificial galaxies. In the lower panels of Fig. \ref{fig:completeness_sex_mto}, we show the R$_e$ and $\bar{\mu}_{e,r'}$ distribution of the LSB galaxies, and indicate the objects detected by each of the methods in the two different panels. 

\indent MTO and SE detected 205 and 176 objects, respectively, of the 244 LSB galaxies. Similarly as was demonstrated with the simulated data, we found also with the real data that MTO was able to increase the completeness limit of the detections toward lower SBs, and both toward larger and smaller effective radii. The completeness limits obtained with the mock images match well those obtained using real detected galaxies in the $\bar{\mu}_{e,r'}$-R$_e$-parameter space.  As can be seen in the lower panels of Fig. \ref{fig:completeness_sex_mto}, most of the missed objects are faint dwarfs with small effective radii. This is expected since we showed (Fig. \ref{figparams_in_out}) that the underestimation of galaxy sizes becomes more severe toward lower SBs. 

\begin{figure}
    \centering
    \includegraphics[width=\columnwidth]{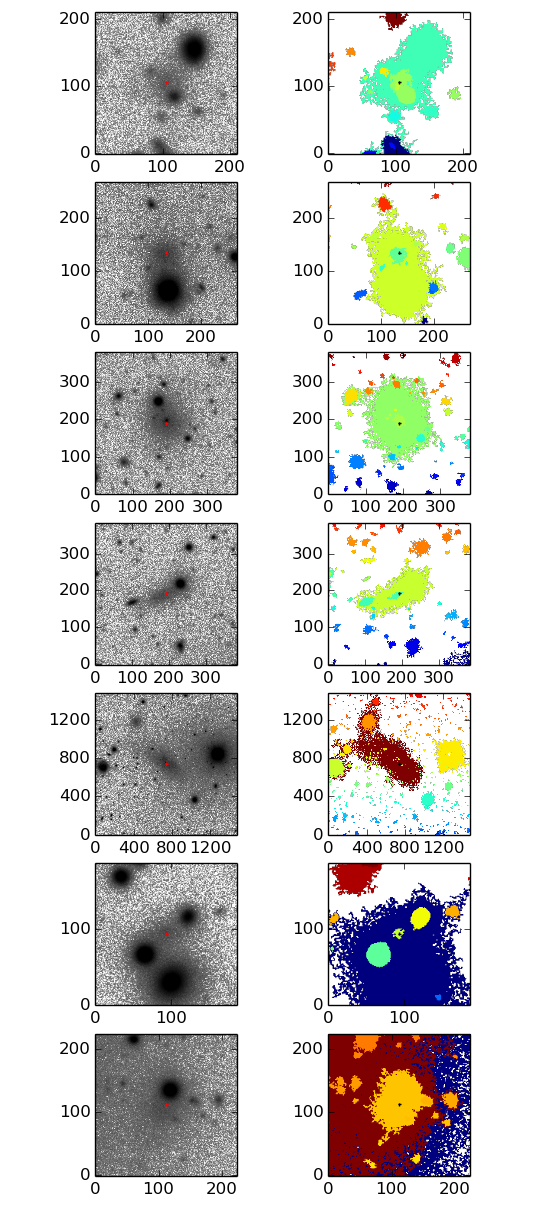}
    \caption{Examples of objects that have relatively bright SBs and large sizes but were not considered as detected with MTO by our test in Sect. 3.2.2 since the center of the detected object passing the $a$ > 2 arcsec condition is too far from the dwarf center. The left panels show the r'-band postage stamps, and the right panels show the cutouts from the segmentation maps. The target objects are shown with the small crosses.}
    \label{fig:missed_by_mto}
\end{figure}

\indent In the bottom right panel of Fig. 1, some (N=8) galaxies with $\bar{\mu}_{e,r}$ < 26 mag arcsec$^{-2}$ and R$_e$ > 4 arcsec are missed by MTO. In principle, such galaxies should be easily detectable in the images. We inspected why they are not identified in our test. In Fig. \ref{fig:missed_by_mto} we show postage stamps of such galaxies and the corresponding segmentation maps. All of those galaxies are detected by MTO, but do not pass our test, as the center of the detection is too much off from the dwarf in the FDS catalogs. This effect is caused by the mixing of the diffuse outskirts of the objects. Based on their size and SB, all these galaxies would have been selected to our catalog (described in Sect. 5). This indicates that the completeness obtained with our test is slightly underestimated.

\indent In summary, our analysis in this and in the previous subsection showed that using MTO we can detect the LSB dwarfs with more than 50 \% completeness down to $\bar{\mu}_{e,r'}$ $\approx$ 26.5 mag arcsec$^{-2}$ in the FDS data, which is approximately one magnitude deeper than with SE. Additionally, MTO pushes the 75\% detection limits near the level of the 50\% completeness limits of the FDSDC. 

\section{Photometry}

In order to obtain reliable photometric measurements for the identified galaxies, we used GALFIT 3.0 \citep{Peng2010} to fit S\'ersic functions to the 2D galaxy light distribution. Since MTO identifies all the structures in the images  very accurately and labels them in the segmentation maps, we were able to automatize the photometric measurements using the information from those maps. The eight steps in the photometric measurements were as follows.

The first step was cutting the postage stamps: Postage stamp images for all the available bands from the science mosaics and weight mosaics were cut. The images were cut according to the central coordinates of objects found by MTO, and cut radii were selected to be ten times the semimajor axis lengths found by MTO. Sigma images were also made accordingly from the weight images following the workflow described by V18.

The second was classifying secondary objects and background: Since the FDS data are deep and the cluster is a dense galactic environment, the postage stamps usually include many galaxies in addition to the main object. Those objects are first selected based on the MTO segmentation maps and their sizes and magnitudes are also collected into a list. Using the segmentation map also the background is identified. As all the pixels in the images belong to some branch of the max-tree (at least the root), the background is selected as the object that has the lowest median SB among the objects that cover at least 20\% of the total image area. The background pixels are then fit with a linear plane and the goodness of fit is estimated by calculating the normalized chi-squared value, $\chi^2_{\nu}$. Based on our experience, we decided to use $\chi^2_{\nu}$<2 as the limiting value for a linear sky model. In the case of nonlinear background, we treat the lowest segmentation branch in the image as a normal object and fit it with a S\'ersic profile.

The third was classifying secondary objects: It is not purposeful to fit all the secondary objects with GALFIT, since the rapidly growing number of fit parameters can make the fitting unstable and many of the secondary objects are small, which means that they can be masked. However, especially in the inner parts of the cluster many objects appear nested on the sky and have to be fitted simultaneously in order to obtain accurate parameters for them. Thus we need to select a limited amount of objects to the GALFIT modeling. At first, we select all the secondary objects that have\footnote{m$_{r',main}$ corresponds to the r'-band total magnitude of the main object.} m$_{r'}$ < m$_{r',main}$ + 3 mag and have total magnitude m$_{r'}$ < 23 mag. From those objects we keep the ones that are bright (m$_{r'}$ < 15 mag), extended ($a$ > 2 arcsec), or overlapping with the main target ($d$ < 3R$_e$). All the remaining objects that are not selected for the model or identified as the background are selected for masking. Additionally, the centers of bright stars (m$_{r'}$ < 15 mag) are selected for masking\footnote{We masked the parts of stars where the SB is brighter than 22 mag arcsec$^{-2}$.}, although their outer parts will be fitted with a S\'ersic profile. At last, in order to keep the number of models in the GALFIT modeling reasonable we limit the maximum number of S\'ersic profiles to five. In cases where there are more than five objects that are initially selected for the model we remove the faintest objects from the model and mask them.

The fourth was creating an input model and a mask: The objects selected for masking are masked based on the MTO segmentation maps. The GALFIT model is initialized as follows: The background is modeled with a linear plane with three degrees of freedom (x- and y-gradients and mean) and all the objects selected for the model are modeled with S\'ersic profiles whose initial magnitude and effective radius are taken from the MTO output lists and axis ratio $b/a$ and S\'ersic index $n$ are set to one. All the parameters are fitted simultaneously, but we apply the following limits for the S\'ersic components in order to avoid divergence in the fitting routine: We limit the center of the S\'ersic profile to be within 3R$_e$ from the detection center, the effective radii to be between 0.1R$_{e,in}$ < R$_{e,out}$ < 10R$_{e,in}$ (R$_{e,in}$ is the initial effective radius), S\'ersic index to be between 0.3 < $n$ < 10, and the magnitude to be within 2.5 mag from the MTO input value.

The fifth was fitting: The input model is fed to GALFIT with the corresponding sigma image and PSF model. We use the PSF models generated by stacking stars described by V18. The parameters of the objects obtained from GALFIT fitting are then collected to a table.

The sixth was identifying the nucleus: Some dwarf galaxies have an unresolved star cluster in their center, which needs to be fitted with a separate $PSF$ component in order to obtain a good fit. We compare the Gini coefficients of the $PSF$ and the galaxy center within the innermost 2 arcsec in order to identify nuclei. The galaxy center is adopted from the S\'ersic model fit performed with GALFIT (step 5) and paraboloid fitting is used in the central regions to adjust the center. If the Gini coefficient of the galaxy center is at least 40\% of that of the $PSF$ step 5 is performed again using S\'ersic + $PSF$ as a input model for the galaxy. More details of the detection of the nucleus are given in Appendix A.5.

The seventh involved aperture magnitudes, the residual flux fraction, and the concentration parameter: Similarly as V18, we also measure aperture magnitudes within R$_e$ using an elliptical aperture with $b/a$ ratio adopted from the GALFIT model. We also calculate the residual flux fraction 
\begin{equation}
RFF\, =\,  \frac{\Sigma_{i=1}^{n_r<R_p} |data_i \, -\, model_i |-0.8\sigma_i}{F_{r<R_p}},
\end{equation}
where $n_r<R_p$ is the number of pixels within the Petrosian radius (R$_p$). R$_p$ is defined to be the radius where galaxy's SB equals 1/5 of the mean SB within that radius. $F_{r<R_p}$ is the total flux within the R$_p$ and $\sigma_i$ is the sigma image value at pixel $i$. The term $|data_i\, -\, model_i |$ corresponds to the residual at a given pixel, and the following term takes into account the expected value for a normal-distributed noise, normalizing the perfect residual to zero. We also calculate the concentration parameter:
\begin{equation}
C\, =\, 5\times\log_{10}\frac{R_{80}}{R_{20}},
\end{equation}
where $R_{80}$ and $R_{20}$ are the radii that enclose 80\% and 20\% of the total Petrosian luminosity of the galaxy, respectively.

The eighth was generating residual and color images: In order to visually classify the identified objects based on their morphology, we also generate color composite images of them using the g', r', and i'-band FDS images. We also generated residual images of each object, by subtracting the GALFIT model from the original image after first transforming each into units of mag arcsec$^{-2}$. Inspecting the residuals in this manner gives a similar weight to residual in each intensity scale and it allows us to easily spot any inner structures such as spiral arms or bars in the galaxies that can give crucial information about the nature of the objects in the classification.

\begin{figure*}
        \includegraphics[width=17cm]{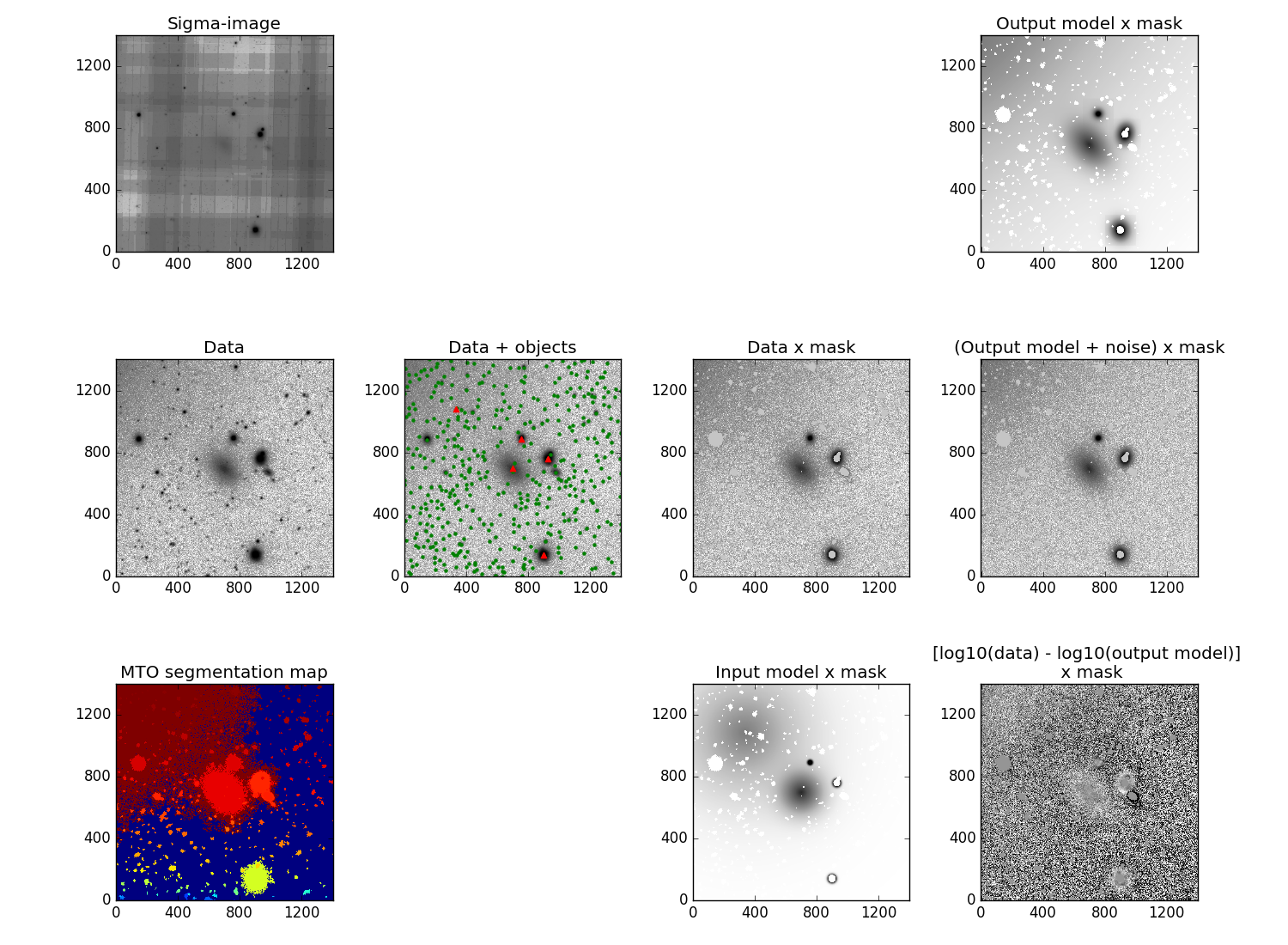}
    \caption{Example of the steps done in order to obtain the photometric parameters. First, the sigma-, science- and segmentation-image postage stamps centered on the object of interest are cut ({\it left panels}). The objects are selected for the mask and the GALFIT model ({\it the green and red symbols in the second column picture from the left, respectively}). A mask is generated ({\it the upper panel of the third column from the left}) using the segmentation map, and the initial model ({\it the lower panel of the third column from the left}) is built using the MTO object locations, magnitudes, and sizes. We note the large fuzzy object in the input model above and to the left of the target, which corresponds to the extended source outside the image field of view: The object is placed within the image in the input model, but it is allowed to move outside the image edges during the fitting. GALFIT is then used for model fitting. The right column panels from top to bottom show the fitted model with and without the noise with the mask overlaid and the residuals after subtracting the model from the data.}
        \label{example_mto_galfit}
\end{figure*}

An example of the different photometric steps is shown in the Fig. \ref{example_mto_galfit}.

\subsection{Accuracy of the photometric pipeline}

\begin{figure*}
        \includegraphics[width=17cm]{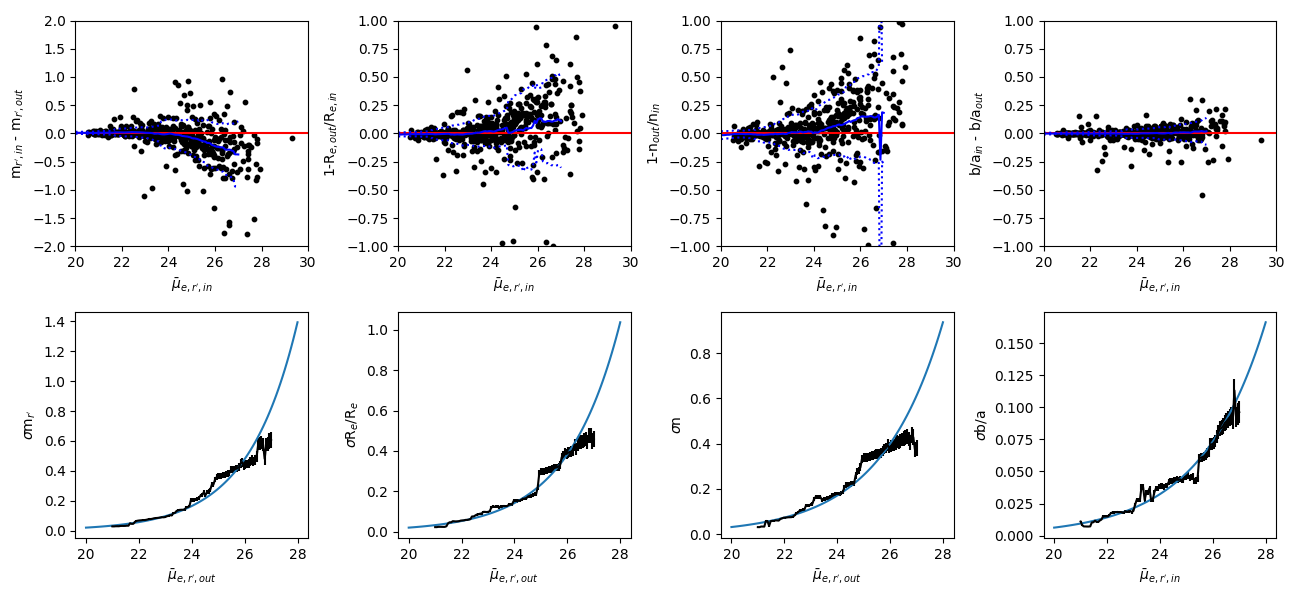}
    \caption{Comparison between the input and output values of the structural parameters of the mock galaxies. The upper panel shows the differences between the input and output magnitudes (m$_{r'}$), effective radii (R$_e$), S\'ersic indices ($n$), and axis ratios ($b/a$) as a function of input mean effective r'-band SB ($\bar{\mu}_{e,r'}$). The black symbols show individual measurements, and the blue line shows the running mean with a 2 mag arcsec$^{-2}$ filter size. The corresponding standard deviations are shown with the dashed blue lines. The lower panels show the magnitude of the standard deviations between the input and output values. The blue curves show the fits of Eq. 4 to the data.}
        \label{fig:photom_accuracy}
\end{figure*}

In order to quantify the uncertainties of the parameters obtained from the photometry, we tested the accuracy of our photometric pipeline with the mock images described in Sect. 2.2. We ran the photometric pipeline for the identified mock galaxies similarly as for real images and compared the obtained parameters of the individual galaxies with the corresponding input values. The comparison between the input and output values as a function of the input SB is shown in the upper panels of Fig. \ref{fig:photom_accuracy}.

\indent In order to estimate the uncertainties in the measured parameters of the real galaxies, we measured the standard deviation, $\sigma$, of the differences between the input and output values of the measured parameters as a function of the output SB. Similarly as \citet{Hoyos2011}, we fit a linear model to the logarithm of the standard deviations:
\begin{equation}
\log_{10}(\sigma)\,=\, \alpha \times \bar{\mu}_{e,r'} + \beta,
\end{equation}
where $\alpha$ and $\beta$ are fitted. The relations between the mean effective SB in r' band and the standard deviations of the difference between the input and output values of the different parameters are shown in the lower panels of Fig. \ref{fig:photom_accuracy}. The fit parameters of the curves in the lower panels are shown in Table \ref{table:fit_params}. 

\indent The scatter between the input and measured values increases with decreasing SB. There are also small systematic trends in the GALFIT parameters with decreasing SB, so the measured magnitudes become slightly fainter and effective radii slightly smaller than the input values, but these trends are much smaller than the scatter. We use these fits for estimating the uncertainties of the parameters in the final catalog. The accuracy  of our pipeline is similar to that obtained by V18, who also used the FDS data, but performed the photometry\footnote{In particular, Venhola et al. created the masks manually and decided on the input model based on visual inspection. These steps could only be reliably automated due to the accurate segmentation of the image by MTO.} manually. This result shows that automated analysis of the LSB galaxies can obtain similar accuracy as the manual one when the accurate segmentation maps of the MTO are used for generating the masks and input models. 

\begin{table}
\caption{Parameters of the fitted curves shown in the lower panels of Fig. \ref{fig:photom_accuracy}.}
\label{table:fit_params}
\centering
\begin{tabular}{c | c | c}
\hline\hline
Parameter & $\alpha$ & $\beta$ \\
\hline
m$_{r'}$ &   0.232 & -6.362\\
R$_e$ &   0.216 & -6.042\\
$n$ & 0.185 & -5.214 \\
$b/a$ & 0.177 & -5.752 \\
\end{tabular}
\end{table}

\section{Preparation of the catalog}

In order to detect LSB galaxies from the full FDS data set, we ran MTO for all the 26 FDS fields. For the detection, we used the same stacked g'+r'+i' band mosaics that were also used for generating the FDSDC in V18. As a result from the detection, we obtained the locations, magnitudes, and the minor- and major-axis radii of the objects. We then used the stellar halo masks generated by V18, which were used to mask the bright stars from the data, and removed all the detections that were within the masked areas.

\indent We then selected the objects with $a$ $>$ 2 arcsec and a median SB of the detected pixels fainter than 23 mag arcsec$^{-2}$. The SB limit guarantees that we mostly detect LSB galaxies that were incompletely detected in the FDSDC. We acknowledge the fact that when using this relatively high SB limit our initial detection lists partly overlap with FDSDC. However, this is not a problem since we subsequently set further limits based on GALFIT parameters, in order to select the objects for the final LSB galaxy catalog, and also remove any duplicate detections of the galaxies. As a result we obtain a sample of 7,270 LSB candidates, which is an order of magnitude more than we currently know.

\indent In addition to galaxies, MTO detects many extended objects that are not galaxies, such as reflection rings around bright stars or artifacts from the image reduction. Additionally, there are background galaxy groups that are located within a common LSB envelope, which gets associated with some of the galaxies, making its median SB low and its size large (see Appendix A.4). We were able to remove most of the reflection rings (see Appendix A.4) by using masks generated by selecting bright stars (m$_{r'}$ < 16 mag) from the American Association of Variable Star Observers’ Photometric All Sky Survey catalog (APASS; \citealp{Henden2012}) and excluding sources, which reside in the areas near them. Around the bright stars, we masked the areas where the SB of the star is $\mu_{r'}$ < 25.5 mag arcsec$^{-2}$. For that task, we used the PSF model of V18. Applying these masks removes $\approx$ 3 \% from the survey area, but as the masks cover areas around saturated and bright stars, those areas would have been non-usable in any case. After excluding the masked objects, we inspected all objects and manually removed all the remaining false positives from the sample (N$_{removed}$=1234).

\indent We then ran the automated photometry for all the remaining galaxies (N=6036) and visually checked all the model fits. The galaxies that were not fitted optimally were flagged. Flagged galaxies were then refitted after modifying their masks and/or input models. Only 41 (=0.7\%) galaxies needed refitting.

\begin{figure*}
        \includegraphics[width=13cm]{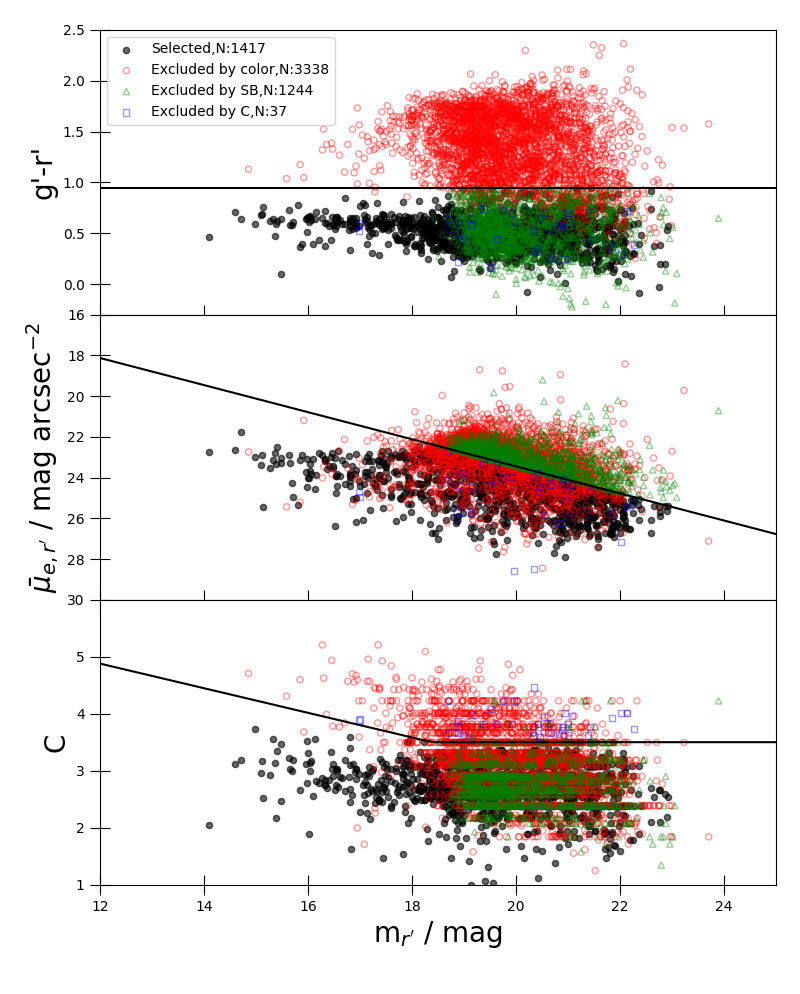}
    \caption{Parametric cuts used for the selection of LSB candidates. The panels show g'-r' colors ({\it top panel}), SBs ({\it middle panel}), and the concentration parameter, $C$ ({\it bottom panel of the galaxies}), as a function of their r'-band apparent magnitude, m$_{r'}$. The selected galaxies ({\it black symbols}) and the galaxies rejected for being too red ({\it red symbols}), having too high of a SB ({\it green symbols}), and a too high $C$ ({\it blue symbols}) are indicated with the different symbols. The black lines in the panels show the selection cuts (see the text and V18). }
        \label{fig:parametric_selection}
\end{figure*}

\indent As our intent is to produce a homogeneous extension to the FDSDC, we followed the steps done by V18 for selecting the likely cluster galaxies. First, we applied the same color, SB, and concentration parameter cuts as Venhola et al. We excluded the galaxies that are redder than the brightest cluster galaxies (g'-r'>0.95 or g'-i'>1.35). We also excluded the small high-SB galaxies whose SB deviates by at least three standard deviations from the mean of the known cluster galaxies' SB distribution (see Fig. \ref{fig:parametric_selection}). Finally, we excluded the very concentrated ($C$>3.5) galaxies with faint apparent magnitude.

\indent After applying the parametric selection, we are left with 1,417 galaxies that pass all the selection criteria. We then crossmatch those galaxies with the galaxies detected previously in the FDSDC. We remove galaxies whose centers are within 5 arcsec from the FDSDC galaxies in order to eliminate the duplicates. This removes 403 galaxies from the sample, leaving 1014 galaxies. The radius of 5 arcsecs radius is adequate as that corresponds to approximately 500 pc at the distance of the Fornax cluster, and the projected separations of the FDSDC galaxies are much larger than that. We also crossmatched our remaining 1014 galaxies with the background galaxies that were identified during the creation of the FDSDC by V18. V18 already classified those galaxies morphologically as background galaxies, so we do not perform that task for the common galaxies again.  We use a search radius of 3 arcsec when searching for duplicate background galaxies, as their sizes are smaller and number density is larger than for the FDSDC galaxies. After removing the 419 duplicates, we are left with 595 previously unidentified LSB candidates.

\indent We inspected the 595 LSB candidates visually using color images generated from the FDS g'-, r'-, and i'-band seeing-selected data\footnote{We used the mosaics generated from seeing selected data by \citet{Cantiello2020} in order to resolve possible spiral and bar structures.}, and residual images, and removed from the sample  the galaxies (N=320) that show a clear spiral structure or a bulge not typically seen in low-mass galaxies. We classified the other galaxies as either likely cluster galaxies (N=226) or galaxies with unclear morphology (N=49). We used the same parametric classifications as V18 for the uncertain cases. We excluded the red (g'-r'>0.45) uncertain objects with high $RFF$ as likely background galaxies (N=10) and accepted the others as likely cluster members (N=39). We did not perform further morphological classifications to early- and late-type galaxies due to the LSB nature of the objects, which makes such  a classification uncertain.

\begin{figure}
        \includegraphics[width=\columnwidth]{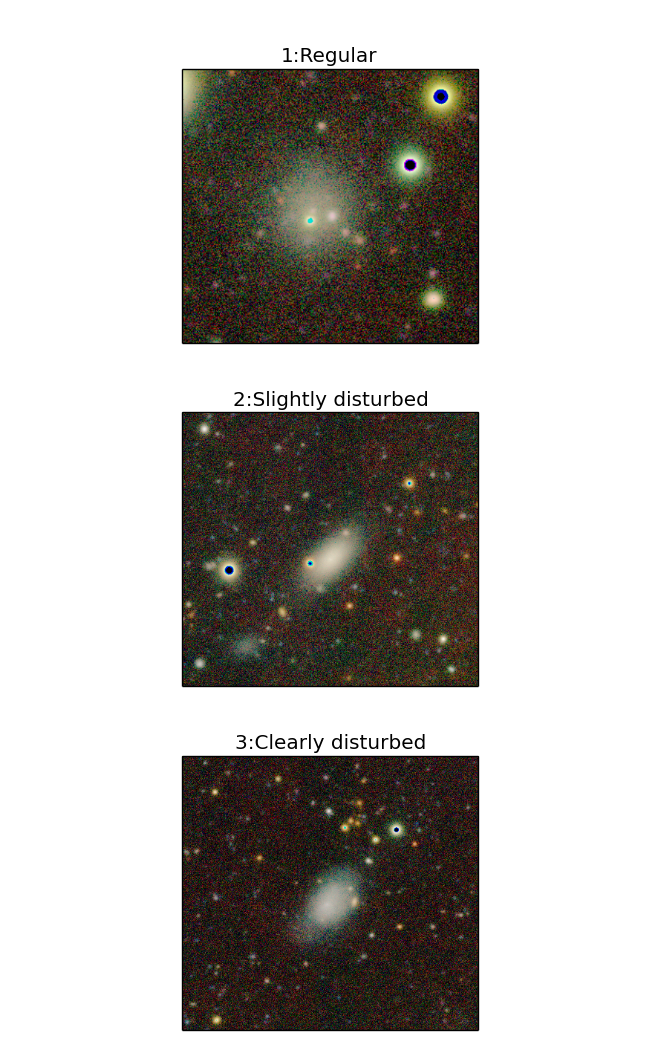}
    \caption{Examples of the different morphological labels given for the galaxies based on visual inspection. The galaxy in the middle panel shows slight irregularities in its outskirts and is thus classified as slightly disturbed. The galaxy in the bottom panel shows clear tidal arms, which justifies its classification as clearly disturbed.}
        \label{fig:tidal_examples}
\end{figure}

\indent Since we aim to use the catalog to analyze the effects of the environment on the evolution of dwarf galaxies, we identified dwarfs whose morphology shows signs of interactions. This classification was done visually (by Aku Venhola) from the images generated by the photometric pipeline. We gave the objects four different classes: (i) ``regular,'' no signs of interactions in the outskirts; (ii) ``possibly disturbed,'' some irregularities in the outskirts; (iii) ``disturbed,'' clear tidal tails or arms in the outskirts; and (iv)  ``unclear,'' some nearby objects or SB of the object prevents the classification.
We did this classification for all the FDSDC galaxies and additional galaxies identified in this work. Examples of the different morphologies are given in Fig. \ref{fig:tidal_examples}.

\indent After all the selection cuts we are left with 265 new LSB dwarfs in the Fornax cluster. When these galaxies are combined with the FDSDC (N=556), we have a total of 821 dwarf galaxies in the area of the FDS. We publish the parameters and classifications of the LSB dwarfs we have found along with those of the FDSDC galaxies as a separate electronic table. An extract of the longer table is presented in the Appendix B.

\section{Analysis of the LSB galaxies}

\subsection{Luminosity function}

\begin{figure}
        \includegraphics[width=\columnwidth]{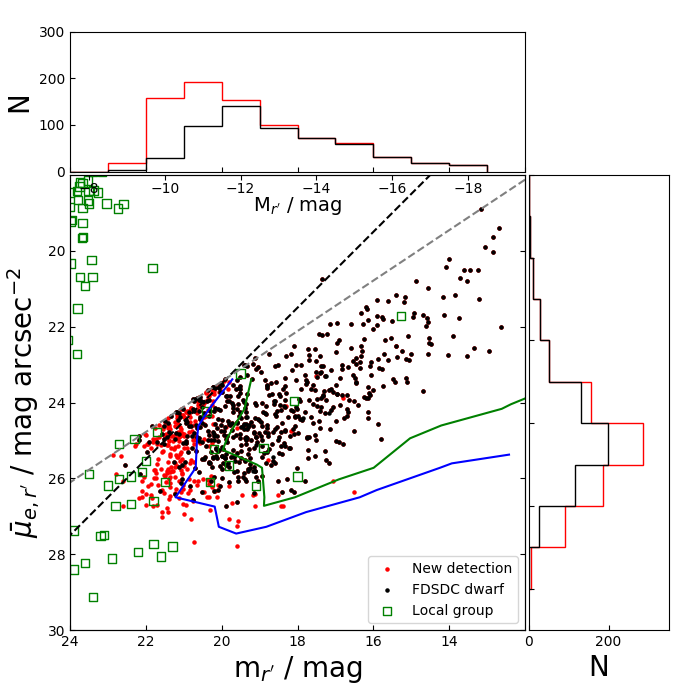}
    \caption{ Apparent r'-band magnitudes (m$_{r'}$) and mean effective SBs ($\bar{\mu}_{e,r'}$) of the FDSDC galaxies ({\it black symbols}) compared with the LSB extension galaxies described in this work ({\it red symbols}). The histograms on the x and y axes show the distributions of the magnitudes and SBs with the same colors as in the scatter plot. The blue and green contours show the 50\% and 75\% detection limits found using the artificial galaxies (Sect. 3.2), and the black and dashed gray lines show the $a$ > 2 arcsec and SB selection limits, respectively. There are also galaxies with R$_e$ < 2 arcsec as the selection was based on MTO detection $a$ > 2 arcsec. For a comparison, we also show a sample of LG dwarfs and globular clusters (green squares clearly above these selection limits) from \citet{Brodie2011}.}
        \label{fig:mag_mu_sample}
\end{figure}

The new LSB galaxies extend the FDSDC toward fainter total magnitudes and lower SBs. Effectively the extension shifts the peak of the detected LF 1 mag deeper than FDSDC. The magnitude and mean effective SB distribution of the LSB extension galaxies are shown with red symbols in Fig. \ref{fig:mag_mu_sample}. As we have defined the completeness limits for our detection algorithm, we can also identify whether the changes in the abundance of dwarfs in the different regions of the parameter space are consequences of the detection incompleteness or real changes in the abundances. Between -17 mag < M$_{r'}$ < -12 mag, the abundance of dwarfs seems to drop both toward low and high SBs. In this region, the abundance drop takes place well before we start to be incomplete in the detections, so it appears to be caused by the span of the intrinsic parameter distribution of dwarfs. On the other hand, for galaxies with M$_{r'}$ > -12 mag we are limited by our completeness and selection limits, which makes it impossible to trace the trends in the dwarf galaxy magnitude-SB distribution in that region.
 
\indent Sizes and luminosities of the Local Group (LG) dwarfs can also give insights whether we are likely to miss many dwarfs. Based on the comparison with LG dwarfs \citep{Brodie2011}, we do not seem to exclude many dwarf galaxies (apart from ultra-faint dwarfs) from our sample with our selection limits if the Fornax cluster and LG dwarf populations resemble each other (see Fig. \ref{fig:mag_mu_sample}). The green symbols clearly above the selection limits of the Fornax galaxies are LG globular clusters.

\begin{figure}
        \includegraphics[width=\columnwidth]{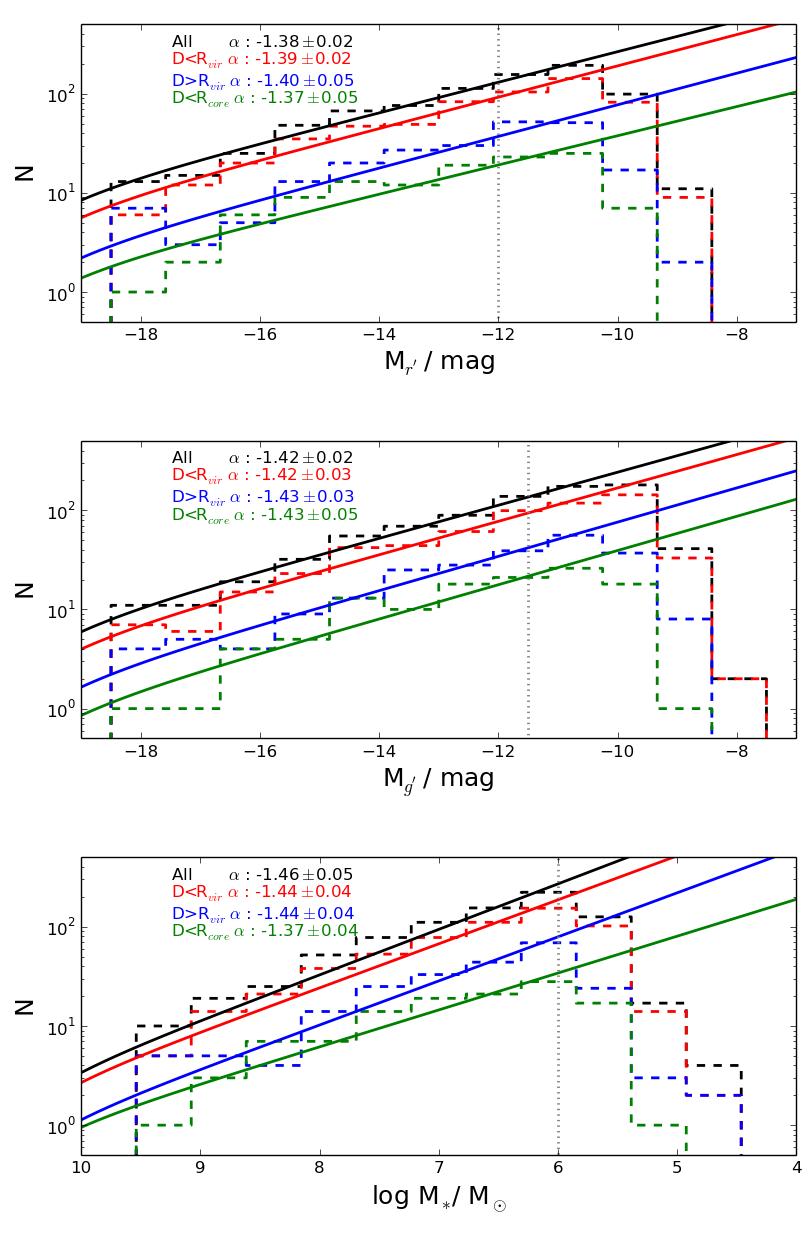}
    \caption{ Upper and middle panels: r'- and g'-band LFs of the Fornax cluster dwarf galaxies, respectively. Bottom panel: Stellar mass function. The black, red, blue, and green histograms show the LF for all the galaxies, for galaxies within the virial radius of the Fornax cluster, those outside of it, and those within the cluster core, respectively. The curves show the Schechter function fits to the data, and the fit parameters are shown with corresponding colors in the upper-left corner of the image. The dotted gray lines show the completeness limits for the different parameters.}
        \label{fig:lf_comparison}
\end{figure}

\indent As this work extends the FDSDC especially in the low-luminosity end, it is interesting to see the effects of these newly identified galaxies on the low-luminosity end slope of the LF. Based on the dwarf galaxy distribution and completeness limits in Fig. \ref{fig:mag_mu_sample}, we can well trace the dwarf galaxy population down to M$_{r'}$ = -12 mag, without any significant completeness corrections for compensating biases arising from selection cuts or detection completeness. In the upper and middle panels of Fig. \ref{fig:lf_comparison}, we show the LF of the dwarf galaxies in the Fornax cluster in r' and g' bands, respectively. For the r' band we used the total magnitudes from S\'ersic fits, and for the g' band we added the g'-r' color measured within the r'-band R$_e$ to those magnitudes. We also show separately the galaxies within and outside of the Fornax virial radius R$_{vir}$ = 2.2 deg (0.7 Mpc). In the bottom panel of Fig. \ref{fig:lf_comparison}, we show the stellar mass function of the Fornax cluster dwarf galaxies. We transform the galaxy luminosities to mass using the transformations of \citet{Taylor2011}, which adjusts the mass-to-light ratio by the galaxy colors (using the colors measured within R$_e$).

\indent We fit the LFs using the Schechter function (\citealp{Schechter1976}):
\begin{equation}
n(M)\mathrm{d}M = 0.4 \ln(10) \phi^*\left[ 10^{0.4(M^*-M)} \right ]^{\alpha+1}\exp\left(-10^{0.4(M^*-M)}  \right ) \mathrm{d}M,
\end{equation}
where $\phi^*$ is a scaling factor, $M^*$ is the characteristic magnitude where the function turns, and $\alpha$ is the parameter describing the steepness of the low-luminosity end of the function. As we are interested in the low-mass slope of the LF, we fit the galaxies in the range -18.5 mag < M$_{r'}$ < -12 mag, and use fixed characteristic magnitude of M$^{*}_{r'/g'}$= -20.48 mag. The M$^*$ was selected based on the analysis of \citet{Smith2009}, who analyzed a large sample of Sloan Digital Sky Survey (SDSS) galaxies.  We fit the Schechter function to the bins in Fig. \ref{fig:lf_comparison} and assume Poisson uncertainty in the bins while doing the fit. We obtain low-mass-end slopes of $\alpha$ = -1.38 $\pm$ 0.02, -1.42 $\pm$ 0.02, and -1.46 $\pm$ 0.05 for r'-band and g'-band LF and stellar mass function, respectively. These values do not vary significantly, when dividing the galaxies into subsamples including galaxies within or outside the virial radius (2.2 deg) of the Fornax cluster. 

\indent The low-mass-end slopes that have been obtained for the luminosity or stellar mass functions previously in the Fornax cluster have large variance. Using the FDSDC galaxies, \citet{Venhola2019} obtained LF slopes between -1.13 < $\alpha$ < -1.44 in different cluster centric bins if no completeness correction was applied. When the LFs were corrected for completeness, the values varied between -1.27 < $\alpha$ < -1.58 in the different bins. For the whole FDSDC the uncorrected slope is $\alpha$ = -1.18, which is significantly flatter than found with this more complete sample of the current work. \citet{Hilker2003} and \citet{Mieske2007} calculated a low-mass slope of $\alpha$ = -1.11$\pm$0.1 from a sample of early-type galaxies in the center of the Fornax cluster. Since they studied smaller sample of galaxies their uncertainties are much larger than ours.  In both abovementioned studies, M$_*$ was fit as a free parameter. In order to test whether that could explain the differences in the low-mass and low-luminosity end of the distribution, we also performed fitting leaving M$_*$ as a free parameter. We then obtained low-mass-end slopes of $\alpha$ = -1.35 $\pm$ 0.03, -1.40 $\pm$ 0.03, and -1.35 $\pm$ 0.02 for the r'-band LF, the g'-band LF, and the stellar mass function, respectively. Hence, the change in M$_*$ does not significantly affect the slope, and the $2\sigma$ difference between our value for $\alpha$ and the value obtained by \citet{Hilker2003} and \citet{Mieske2007} is not caused by the different fitting technique.

\subsection{Intrinsic shapes and surface brightness correlations}

As seen in Fig. \ref{fig:mag_mu_sample}, at any given luminosity dwarf galaxies have a range of different mean effective SBs. Additionally, there is a trend between the SBs and absolute magnitudes of dwarfs. Here we investigate, how much of this variation in the SBs at given mass or luminosity is due to the projection effects caused by the inclination of galaxy and different stellar populations of dwarfs, and whether the remaining differences are connected to the galactic environment.

\indent Assuming a dwarf galaxy is shaped like an oblate spheroidal and the dust does not obscure the stellar light significantly, its apparent SB is highest when it is viewed edge-on, and lowest when viewed face-on. For a prolate (cigar-like) shape, the apparent SB is the highest when viewed along the longest axis. When the shape is triaxial, the relation between the viewing angle and the SB is more complicated. In order to reduce the bias in the observed SBs of the galaxies, this projection effect can be corrected if the shapes of a galaxy population are known. 

\begin{figure*}
    \includegraphics[width=14cm]{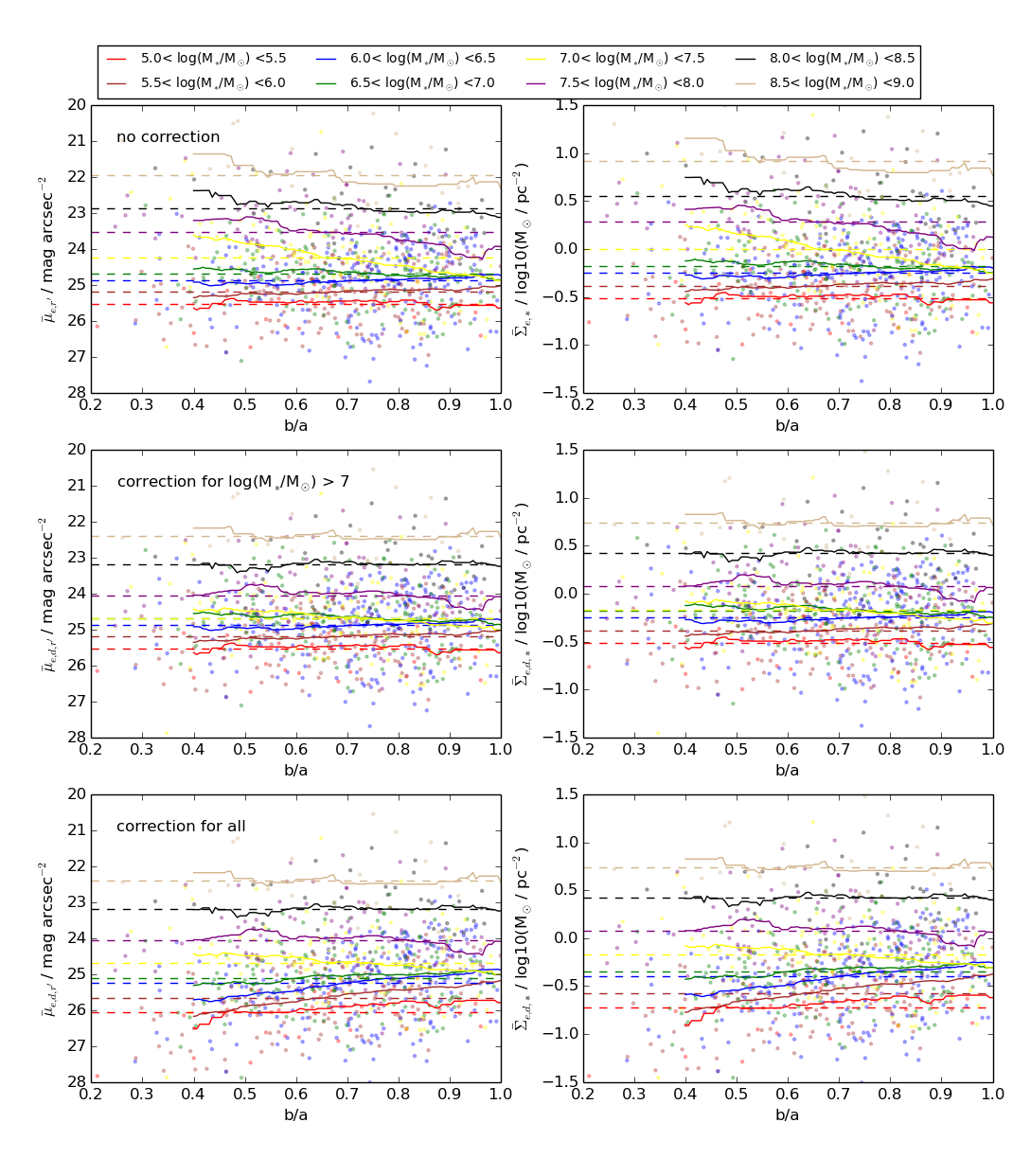}
    \caption{Correlation between the axis ratio ($b/a$), mean effective SB ($\bar{\mu}_{e,r'}$), and stellar SD ($\bar{\Sigma}_{e,*}$) of the galaxies. The upper panels show the values without any inclination corrections, in the middle panels galaxies with log(M$_*$/M$_\odot$) > 7 have been corrected for their inclination, and in the bottom panels inclination correction has been applied to all the galaxies. The different colors correspond to different mass bins, as shown in the legend. The solid and dashed lines indicate the running mean (filter size of 0.2) and the mean of the mass bin. }
        \label{fig:inclination_correction}
\end{figure*}

\indent In order to test whether significant projection effects exist among the dwarfs, we show the correlation between $b/a$ and $\bar{\mu}_{e,r'}$ and mean effective stellar SD ($\bar{\Sigma}_{e,*}$), in the upper panels of Fig. \ref{fig:inclination_correction} for dwarfs in different stellar mass bins\footnote{We calculated the stellar surface densities as in \citet{Trujillo2020}. }. We find that for galaxies with M$_*$ > 10$^7$ M$_\odot$, the SBs and surface densities increase remarkably with decreasing axis ratio. This is consistent with the behavior of disks. For galaxies less massive than that, there is no correlation between the SD or brightness of the galaxy with its apparent axis ratio.

\indent In order to correct for the projection effects, we calculate the de-projected mean effective SB for each galaxy. For a thick oblate disk with a S\'ersic-like luminosity profile, such as the massive dwarf galaxies in the Fornax are, the de-projection can be simply made by multiplying the apparent SB (in flux) by the apparent axis ratio (Salo et al., {\it in prep.}). Thus, we define de-projected mean effective SB $\bar{\mu}_{e,d}$ = $\bar{\mu}_e$ - 2.5$\log_{10} \left( b/a \right)$ for dwarfs with M$_*$ > 10$^7$ M$_\odot$, and $\bar{\mu}_{e,d}$ = $\bar{\mu}_e$ for dwarfs with M$_*$ < 10$^7$ M$_\odot$. We show the relation between the de-projected SBs and densities with the axis ratios of the galaxies in the lower panels of Fig. \ref{fig:inclination_correction}. The correlations with the axis ratios disappear when de-projected quantities are used. We use the de-projected SBs in this subsection. In the bottom panels of Fig \ref{fig:inclination_correction}, we demonstrate that if the same inclination corrections that are used for the M$_*$ > 10$^7$ M$_\odot$ galaxies are applied for lower-mass galaxies, this introduces an artificial trend between $\bar{\mu}_{e,d}$ and $b/a$, which indicates that the assumption about disk shape for those galaxies is wrong.

\begin{figure}
        \includegraphics[width=\columnwidth]{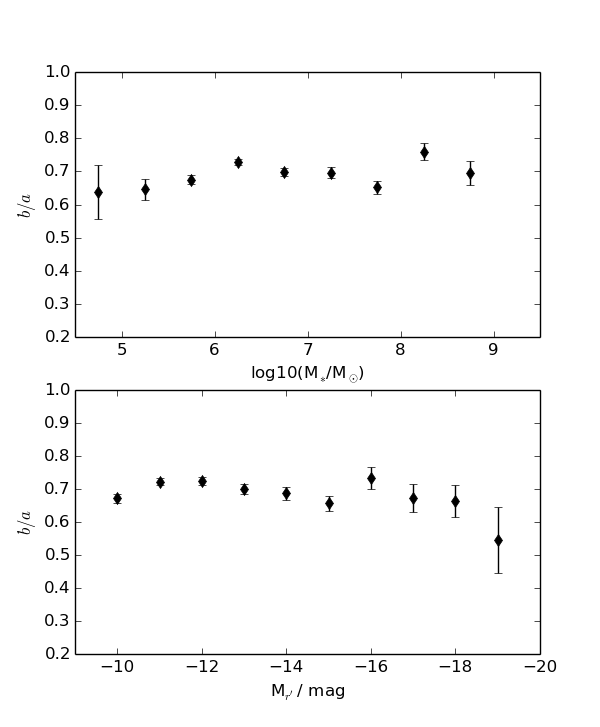}
    \caption{Average axis ratios of the dwarf galaxies in the Fornax cluster in mass (upper panel) and absolute r'-band magnitude (lower panel) bins. The symbols indicate the average values with their uncertainties. The mass bins have a width of 0.6 dex, and the magnitude bins have a width of 1 mag. } 
        \label{fig:ba_in_mass_bins}
\end{figure}

\indent In principle, this result could be caused by a selection bias: If we are biased toward selecting edge-on galaxies in the very low-luminosity end, this could diminish the correlation between the SB and axis ratio when analyzed in mass bins. However, according to our tests made with the mock galaxies, this seems not to be the case (see Fig. \ref{fig:completeness_sex_mto}) as there is no selection bias related to the axis ratio in the low-mass end. In order to confirm that using the Fornax cluster dwarf galaxies, we show the $b/a$ distribution in mass and magnitude bins in Fig. \ref{fig:ba_in_mass_bins}. It appears that the axis-ratio distribution is nearly constant in the low-mass bins, and that most of the low-mass galaxies have $b/a$ $\approx$ 0.7. 

\begin{figure}
    \centering
    \includegraphics[width=\columnwidth]{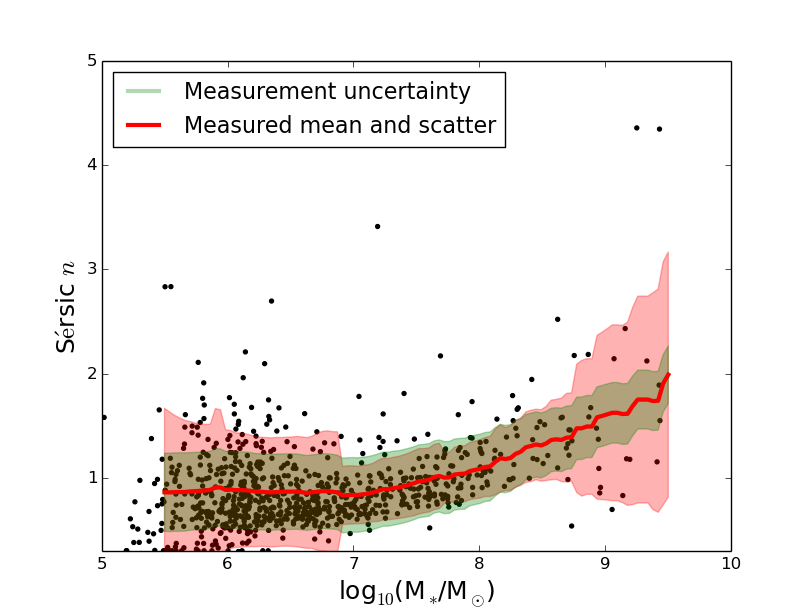}
    \caption{S\'ersic indices as a function of galaxy stellar mass. The black points show the measurements for the dwarf galaxies in the Fornax cluster, and the red line shows the running average with a filter size of 0.5 dex. The green and red shaded areas correspond to the scatter expected from the measurement uncertainties and the measured scatter, respectively.}
    \label{fig:n_vs_mass}
\end{figure}

\indent  In addition to the geometry of dwarfs, we analyze the concentration of their light profiles as a function of their mass. Typically disks of disk galaxies are well fit with an exponential profile, which corresponds to $n$ = 1, and elliptical galaxies and the central bulges of disk galaxies have n $\geq$ 2. In Fig. \ref{fig:n_vs_mass}, we show the S\'ersic indices of galaxies as a function of their stellar mass. The running average shows that low-mass galaxies have on average $n$ $\approx$ 1 or slightly below that and the concentration of the profiles increases toward more massive galaxies. In order to find out whether the scatter of $n$ values around the average is due to the measurement uncertainties or intrinsic to the dwarf population, we compare the measured scatter of the population with the measurement uncertainties that we estimated using the mock galaxies in Sect. 4.1. We find that the measured scatter ({\it red highlighted area} in Fig. \ref{fig:n_vs_mass}) in the S\'ersic indices is similar to the measurement uncertainty {\it green highlighted area} in Fig. \ref{fig:n_vs_mass}) for galaxies with $\log_{10}$(M$_*$/M$_{\odot}$) = 7--8.5. For galaxies more massive than that, the dwarf population seems to have intrinsic variation in $n$ much larger than the measured uncertainties. Galaxies  with $\log_{10}$(M$_*$/M$_{\odot}$) < 7 seem to have slightly larger scatter than expected from the measurement uncertainties. To summarize, the light profiles of most low mass dwarf galaxies are well approximated by an exponential profile and it is possible that all the measured deviations from that are due to measurement uncertainties. Massive dwarfs however have much higher scatter in $n$ values than the measurement uncertainties, which indicates an intrinsic differences in the profiles of the dwarf galaxies.

\indent In order to investigate the connection between a galaxy's SB and its other properties, we calculated the running mean along the log($M_*$)-$\bar{\mu}_{e,d,r'}$ - relation of the galaxies, and calculated the running standard deviation of the galaxies with respect to that trend. Similarly as was done using the SB of the galaxies, we also calculated the means using surface mass densities. We transformed the de-projected mean effective SBs into the corresponding surface densities ($\bar{\Sigma}_{e,d,*}$).

\begin{figure*}
        \includegraphics[width=17cm]{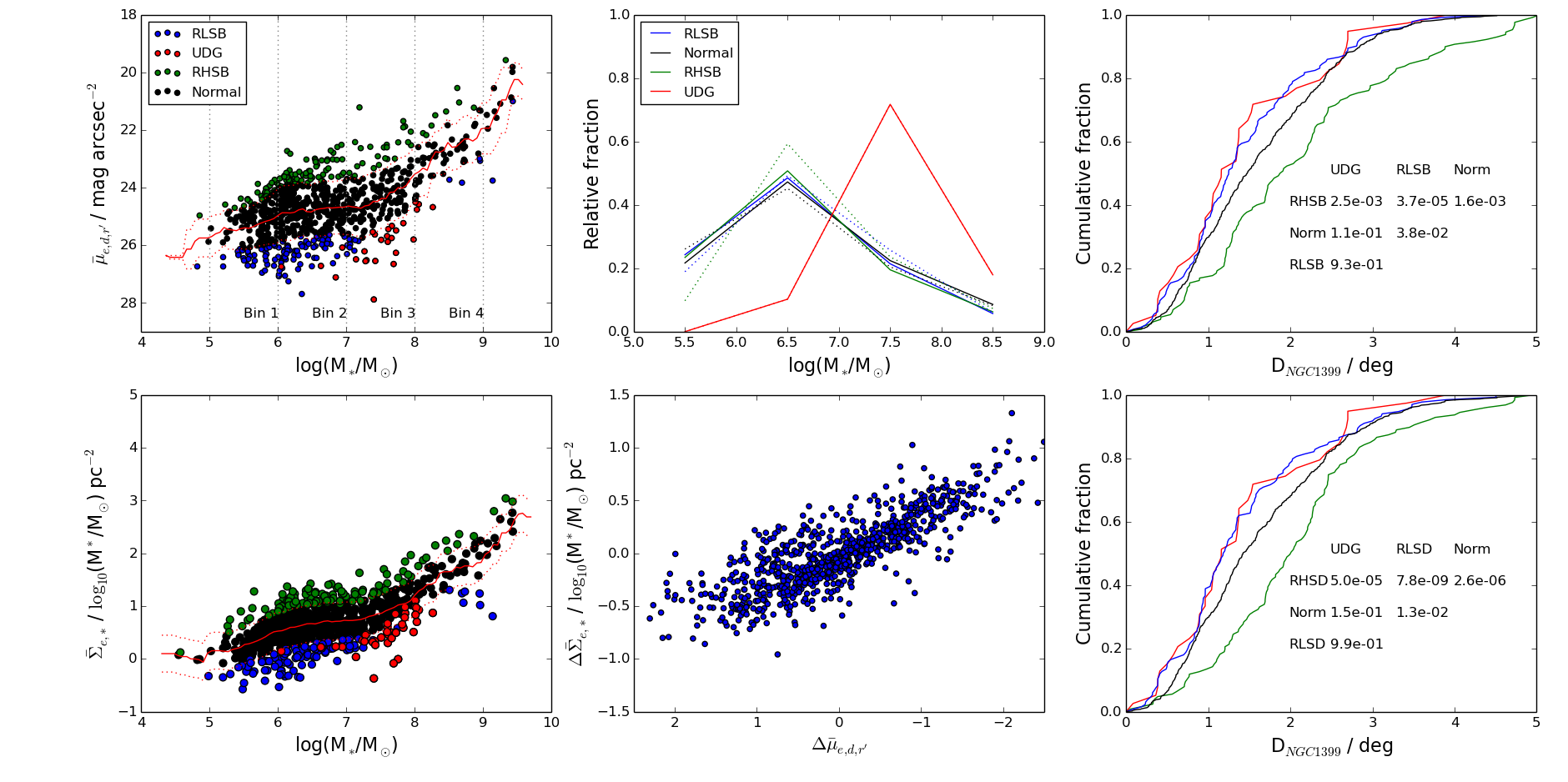}
    \caption{ Upper-left and lower-left panels: De-projected mean effective SB ($\bar{\mu}_{e,d,r'}$) and stellar SD ($\bar{\Sigma}_{e,d,*}$) as a function of galaxy stellar mass (M$_*$) for the Fornax cluster dwarfs, respectively. The solid and dashed red lines show the running mean and standard deviations from that, respectively. The RHSB galaxies, RLSB galaxies, normal galaxies, and UDGs are indicated with the different colors, shown in the legend in the upper-left panel. The definitions of those classes are described in the text. The upper-middle panel shows the normalized mass distributions of the different SB subsamples in the mass bins indicated with the vertical lines in the upper-left panel. The lower-middle panel shows the correlation between $\Delta\bar{\mu}_{e,d,r'}$ and $\Delta\bar{\Sigma}_{e,d,*}$. The right panels show the cumulative cluster-centric distance distribution for the different subsamples when defined using $\bar{\mu}_{e,d,r'}$ (upper-right panel) and $\bar{\Sigma}_{e,d,*}$ (lower-right panel). K-S test p values for the assumption of two samples being drawn from the same distribution are also listed for all the sample pairs in the lower right of the right panels.}
        \label{fig:dist_in_clust}
\end{figure*}

\indent The relation between $\bar{\Sigma}_{e,d,*}$, $\bar{\mu}_{e,d,r'}$ and M$_*$ of the galaxies are shown in Fig. \ref{fig:dist_in_clust}. Trends in the SBs and surface densities are very similar, and have similar spread when the factor of 2.5 between the two quantities is taken in account. The deviations from the means of these two quantities are strongly correlated as can be seen in the lower middle panel of Fig. \ref{fig:dist_in_clust}. In other words this means that, on average, the galaxies that have brighter or fainter SB than the mean SB at given mass, have also SD that deviates similarly from the average. Thus, the difference in the SB is mostly due to an actual variation in stellar mass density, not due to projection or stellar population effects 

\indent In order to neutralize the effects of mass and study how the environment affects the SB of galaxies, we then divided the galaxies into three parts: Relatively high surface brightness (RHSB) galaxies are those that have a SB  > 1$\sigma$ from the running mean with respect to stellar mass; relatively low surface brightness (RLSB) galaxies  deviate by > 1$\sigma$ toward fainter SBs; and the galaxies in between those groups are ``normal'' galaxies. In addition, we studied a subgroup of LSB galaxies, UDGs, defined as galaxies having $\bar{\mu}_{e,d,r'}$ > 24 mag arcsec$^{-2}$ and R$_e$ > 1.5 kpc. Similarly, we separated galaxies based on their SD into relatively low surface density (RLSD) and relatively high surface density (RHSD) groups. The different classes are shown in the upper-left panel of Fig. \ref{fig:dist_in_clust}. The normalized M$_*$ distributions of the different SB groups are also shown in the upper-middle panel of Fig. \ref{fig:dist_in_clust}. 

\indent  We calculated the cumulative distributions of the galaxies' cluster-centric distances (projected distances to NGC 1399) for the different galaxy subgroups. These cumulative profiles are shown in the right panels of Fig. \ref{fig:dist_in_clust}. It appears that the RHSB  and RHSD galaxies are less centrally concentrated in the cluster than the diffuse ones. We calculated also the Kolmogorov-Smirnov (K-S) test p values for the assumption of any two of the subgroups being drawn from the same underlying distribution. The K-S test results between the subgroups are shown in the right panels of Fig. \ref{fig:dist_in_clust}. We find that significant differences are found between the UDGs and RHSB galaxies and between the UDGs and RHSD galaxies (p values 2.5$\times10^{-3}$/5.0$\times10^{-5}$), RLSB galaxies with respect to RHSB galaxies (p=3.7$\times10^{-5}$), between normal galaxies and RHSB, and between normal and RHSD galaxies (p=1.6$\times10^{-3}$/2.6$\times10^{-6}$). When using surface densities, the difference between the RLSD and normal galaxies is also significant (p=0.013).

 \indent As the  galaxies with different surface densities are distributed differently in the cluster, the galactic environment is likely to be responsible for those differences. In  order to further understand how RHSD and RLSD differ from each other, we calculated normalized deviations from mean for a range of parameters and compared those with the excess of galaxy's SD with respect to the mean at given mass. In practice, the normalized deviations were calculated by deriving a running average of the parameter at the given stellar mass, and then normalizing the deviations by running standard deviations.

\begin{figure}
        \includegraphics[width=\columnwidth]{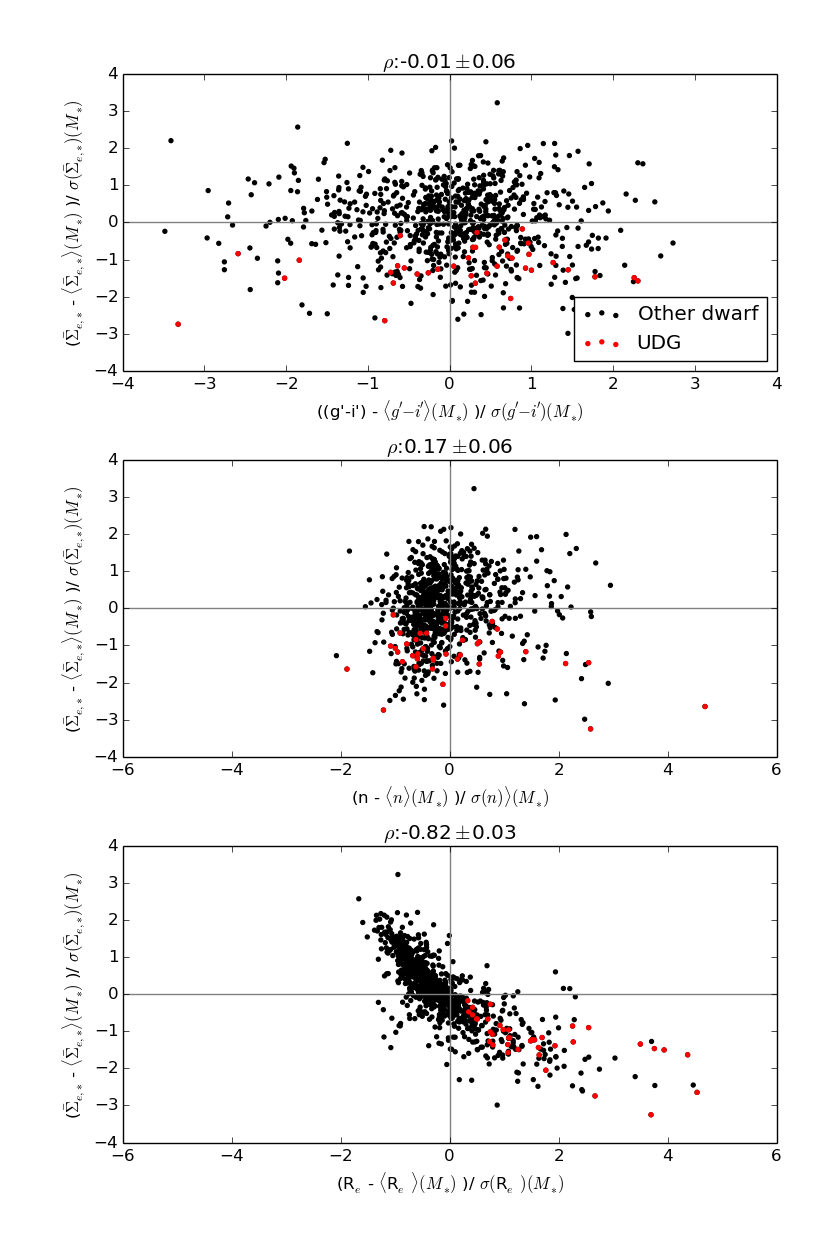}
    \caption{ Correlations between the excess mean effective stellar SD and properties of the FDS dwarfs. The upper panel shows the comparison with the excess color (with respect to the mean at a given mass), the middle panel with the excess S\'ersic index, and the lower panel with the excess effective radius. The red symbols highlight the UDGs, and the black symbols show the other dwarfs. }
        \label{fig:sfd_correlation}
\end{figure}

\indent In Fig. \ref{fig:sfd_correlation} we compare galaxies' normalized deviations from the mean g'-i' color, the effective radius, and S\'ersic index with their deviations from the mean SD. The deviations are normalized by the running standard deviation calculated within the same bin as the running average. We used g'-i' color, as it is available for the whole sample and is independent from the r' band, which is used to calculate the SB and R$_e$. We find that galaxies' SD has no correlation with their color and only a slight correlation with their S\'ersic indices, so dense galaxies are slightly more centrally peaked than the average. As expected, the clearest correlation between the galaxy density is with the galaxies' effective radii, so the higher the galaxy's SD is the smaller R$_e$ it has. The correlation between $\bar{\Sigma}_{e,d,*}$ and R$_e$ is not fully linear due to the differences in stellar populations. 

\section{Discussion}

The reanalysis of the FDS images has improved the completeness of the FDSDC in the LSB regime, which makes it interesting to compare our findings with other deep dwarf galaxy surveys and the theoretical framework.

\subsection{Environmental dependence of the low-mass end of the luminosity function}

There are some works on the dwarf galaxy LF that match our image depth and the detection completeness. In nearby groups, there is the advantage that galaxies can be resolved into stars, which makes it possible to be complete down to fainter magnitudes. Thus, we first make a comparison with nearby group environments. 

\begin{figure}
        \includegraphics[width=\columnwidth]{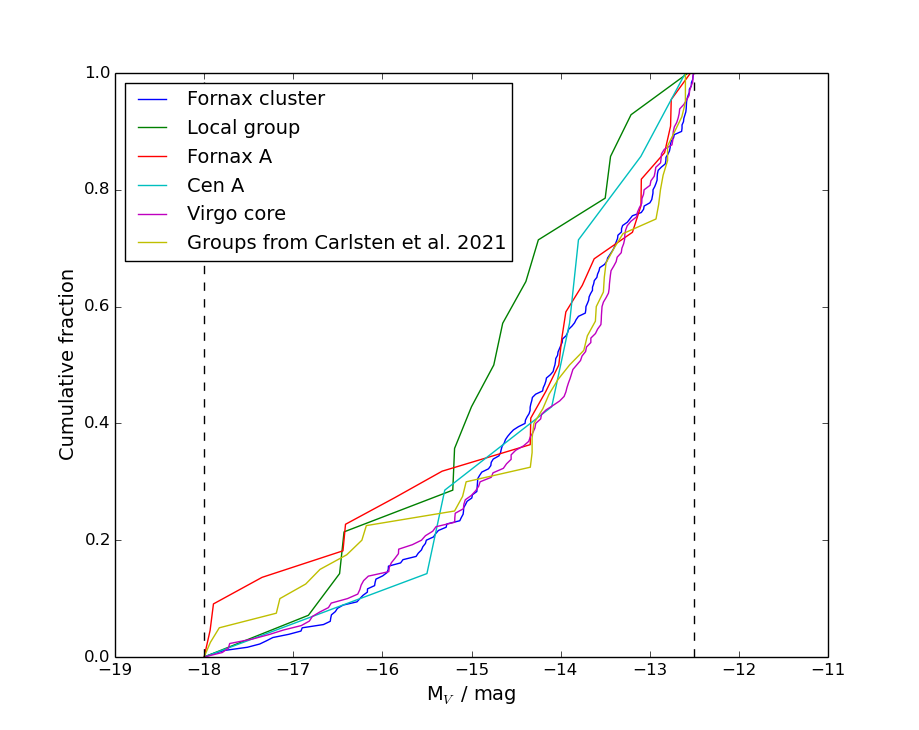}
    \caption{ Comparison of the V-band LFs in the Fornax cluster, the LG, and other nearby environments. We use the compilation of \citet{Ferrarese2016} for the LG and Virgo, the \citet{Muller2019} compilation for the Centaurus A group, the compilation of \citet{Carlsten2020}  for other nearby groups, and the galaxies in this work for the Fornax cluster and the Fornax A group. The vertical dashed lines show the cutoff limits of M$_V$ = -18 mag and M$_V$ = -12.5 mag. } 
        \label{fig:lf_LG_For}
\end{figure}

\indent Low-mass dwarf galaxies have been well studied in the LG. We used the compilation of LG galaxies made by \citet{Ferrarese2016} to compare the LFs in the LG and in the central parts of the Fornax cluster (D < R$_{200}$). We first transformed our magnitudes into V-band magnitudes using the transformation formula\footnote{The formula used is derived from Table 3 of Jordi et al. (2006): V-g   =     (-0.565 ± 0.001)$\times$(g-r) - (0.016 ± 0.001). The transformation was done for galaxies with M$_V$ > -18 mag.} of \citet{Jordi2006}; we compare the cumulative LFs of these two environments in Fig. \ref{fig:lf_LG_For}. We made the comparison for galaxies with  M$_V$ = -12.5 mag since both samples are more than 75\% complete at that magnitude limit. We find that the low-mass end of the LF is steeper in the Fornax cluster than in the LG, but the number of galaxies in the LG is not large enough to exclude the possibility that these distributions follow the same underlying distribution (p=0.28). 

\indent In order to improve the statistics of the group environment LF and thus obtain more insight about possible differences, we add also other groups to our comparison. \citet{Carlsten2020,Carlsten2019} gathered a compilation of satellite galaxy LFs in nearby groups with deep observations. We add the satellite LFs of NGC 1023, NGC 2903, NGC 4258, NGC 4565, NGC 4631, M 51, M 104, and M 81 from that work into our comparison. Since many of the groups of that work contain few satellites, we combine the LFs of those groups as a single group LF. In addition, to those groups we add the LF of Centaurus A by \citet{Muller2019} into our comparison, and show also the LF of the Fornax A subgroup separately, in order to study whether there are differences between the LFs in the different environments. These LFs are shown in Fig. \ref{fig:lf_LG_For}.

\indent Another well-studied cluster environment is the center of the Virgo cluster. \citet{Ferrarese2016} studied the core of the cluster using the data of the Next Generation Virgo Cluster Survey (NGVS, \citealp{Ferrarese2012}, \citealp{Ferrarese2020}), which has similar depth as the FDS. The NGVS is complete down to M$_g$ = -12.5 mag. 

\indent All the abovementioned satellite compilations include a lot of galaxies even with M$_V$ > -10 mag, but the completeness drops significantly in the low-luminosity end. Therefore, we adopted M$_v$ = -12.5 mag for the lower-luminosity limit. In this region, all of the  considered galaxy samples should be fully complete. 

\indent In Fig. \ref{fig:lf_LG_For} we show the cumulative LFs of the different environments. There seems to be differences between some of the cumulative LFs of the different environments. In order to quantify whether the LFs have statistically significant differences, we use K-S tests between the different samples. We tabulate the values in Table \ref{tab:ks-lum}. Based on the K-S tests, we find no statistical evidence of differences between the different samples.

\indent These comparisons show that the different deep surveys show similar behavior of the faint end of the LF down to M$_V$ = -12.5 mag. These results seem robust in light of the available data. For the galaxies fainter than that magnitude limit, it is likely that some of the galaxies will be missed as their $\bar{\mu}_{e,r'}$  drops below 27 mag arcsec$^{-2}$ and some of the galaxies will be unresolved with the typical ground-based seeing of 1 arcsec. Obtaining certainty about the universality of the LF and extending the results toward lower luminosities would require a homogeneous ultra-deep survey with good seeing that has a large sky coverage and thus a large range of different environments. Future surveys such as Euclid (\url{https://www.euclid-ec.org/}) and the Legacy Survey of Space and Time (LSST; \citealp{Ivezic2019}) will likely shed light on this issue. 

\begin{table}
\caption{Results of the K-S tests (p values) when comparing the likelihood that the V-band LFs of different samples are drawn from the same distribution. Fornax and Fornax A data are from this work, Centaurus A data are from \citet{Muller2019}, LG and Virgo data are from \citet{Ferrarese2016} and \citet{Ferrarese2020}, respectively, and the data for other groups are from \citet{Carlsten2020}. The last column lists the number of galaxies in the samples.}

\begin{tiny}

\begin{tabular}{|l|c|c|c|c|c|c|}
\hline
 &  Fornax & Fornax A & LG & Virgo & Groups & \#Galaxies \\
\hline\hline
CenA   & 0.92 & 0.75 & 0.27 & 0.88 & 0.88 & 7 \\
Groups & 0.63 & 0.94 & 0.13 & 0.80 &      & 40 \\
Virgo  & 0.35 & 0.79 & 0.13 &      &      & 130 \\
LG     & 0.28 & 0.27 &      &      &      & 14 \\
ForA   & 0.81 &      &      &      &      & 22  \\
Fornax &      &      &      &      &      & 180  \\
\hline
\end{tabular}
\end{tiny}
\label{tab:ks-lum}
\end{table}

\subsection{Comparison of the luminosity function with IllustrisTNG }

One important test for cosmological simulations is that they need to produce the right amount of luminous low-mass galaxies. The LF obtained from FDS data can be used for making such a comparison. We perform the comparison with the recent IllustrisTNG \citep{Pillepich2018} cosmological simulation, which includes prescriptions for baryonic matter physics. 

\indent IllustrisTNG is a set of full-scale cosmological hydrodynamical simulations that include dark matter, gas, stars, stellar winds, magnetic fields, and black hole components. We analyzed its 110.7$^3$ Mpc$^3$ simulation box (TNG100), which has baryonic and DM mass resolutions of 1.4$\times10^6$ M$_\odot$ and 7.5$\times10^6$ M$_\odot$, respectively, and softening lengths of 185 pc for gas and 740 pc for DM and stellar particles. Since the softening lengths are large compared to the typical sizes of the Fornax dwarfs (R$_e$ $\approx$ 700 pc), it is not safe to compare the structure of the observed and simulated dwarfs. Thus, we restricted our comparisons to the stellar mass function of galaxies.

\indent The TNG100 simulation includes 22 galaxy clusters within the mass range 5-10$\times 10^{13}$ M$_{\odot}$, similar to the Fornax cluster, 7$\times 10^{13}$ M$_{\odot}$ \citep{Drinkwater2001}. We made comparisons with the LF of all these clusters but also compared the Fornax cluster specifically with group number 21 of the IllustrisTNG-100 since it is the most Fornax-like cluster from the group of 22. Cluster 21 has a mass (7.34$\times$10$^{13}$ M$_{\odot}$), a size (R$_{200}$ = 0.56 Mpc), and a 2D velocity dispersion ($\sigma$ = 318 km s$^{-2}$) that very closely  resemble those of the Fornax cluster (M$_{200}$=7$\times$10$^{13}$ M$_{\odot}$, R$_{200}$=0.7 Mpc, and $\sigma$=370 km s$^{-1}$).

\begin{figure}
        \includegraphics[width=\columnwidth]{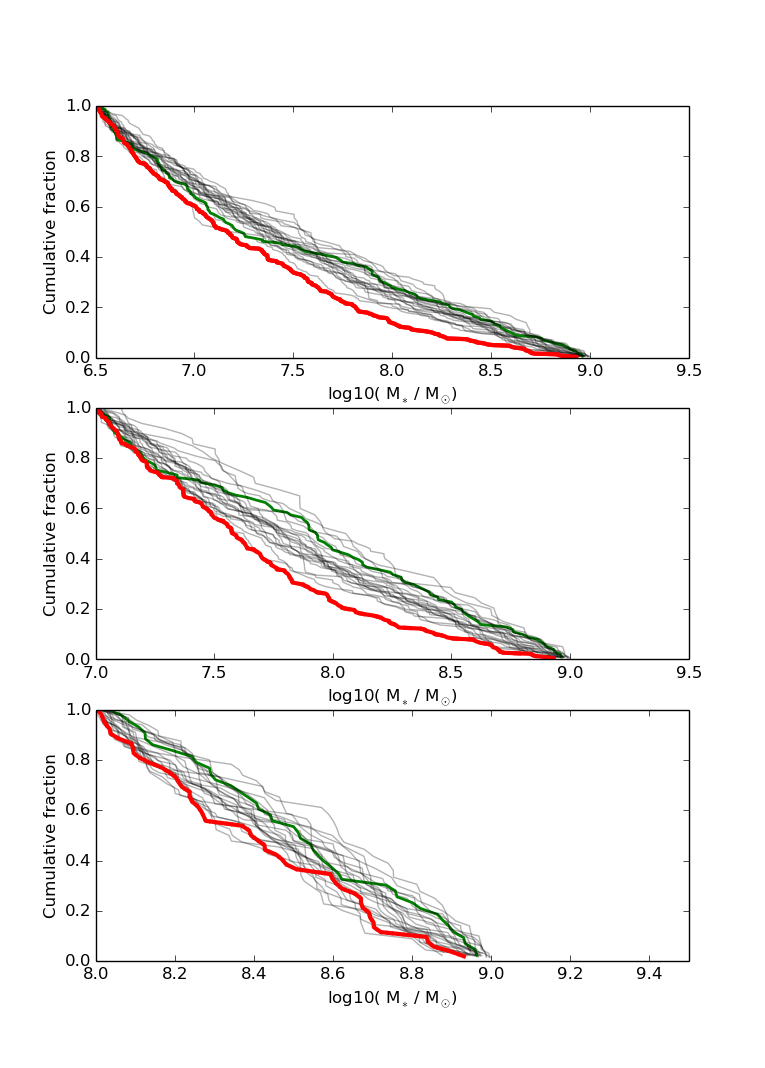}
    \caption{ Cumulative stellar-mass functions for the Fornax cluster (red lines) and IllustrisTNG Fornax-like groups (gray lines). The IllustrisTNG group most similar to Fornax is shown with the green lines. The different panels show the comparisons using different stellar-mass ranges: The upper panel shows galaxies with 6.5 < $\log$(M$_*$/M$_\odot$) < 9, the middle panel shows galaxies between 7 < $\log$(M$_*$/M$_\odot$) < 9, and the bottom panel shows galaxies with 8 < $\log$(M$_*$/M$_\odot$) < 9.} 
        \label{fig:illustris_comp}
\end{figure}

\indent We show the comparisons between the stellar mass functions of the 22 IllustrisTNG clusters and the Fornax cluster in Fig. \ref{fig:illustris_comp}. We performed these comparisons down to $\log$(M$_*$/M$_\odot$) > 6.5, which corresponds roughly to the 75\% completeness of our catalog. We compare the stellar mass functions between 6.5 < $\log$(M$_*$/M$_\odot$) < 9, 7 < $\log$(M$_*$/M$_\odot$) < 9,  and in the range where IllustrisTNG resolves the galaxies well, $\log$(M$_*$/M$_\odot$) > 8. We find that, generally, IllustrisTNG has a shallower low-mass-end slope than that found by us in the Fornax cluster within all the selected limits. These differences are statistically (>2$\sigma$) significant for 18, 17, and 1 clusters when the low, mid, and high lower-mass limits are used, respectively. For the most Fornax-like cluster, we obtained K-S test p values of 0.001, 0.0009, and 0.4 for the low, mid, and high lower-mass limits.  Clearly, the differences become smaller and stop being statistically significant when the comparison is made only for the high-mass range, where galaxies are well resolved by the simulation. On the other hand, there are only few galaxies in the highest-mass range, so the statistical strength of the test is not very good in that part.

\indent A similar conclusion to ours regarding the low-luminosity-end LF slopes between Cen A and IllustrisTNG-100 was also obtained by \citet{Muller2019}. In their comparison, which was made for groups with lower masses than those in this work, they found that the observed LF low-luminosity end is steeper than found in the simulation.

\subsection{Shape of the low-mass galaxies}

We found that galaxies with log$_{10}$(M$_*$/M$_{\odot}$) > 7  require an inclination correction in order to remove the correlation between the SB and apparent axis ratio of the galaxies. On the other hand, the less massive galaxies do not have a correlation between these observables, which can indicate that they are not (thin) disks or that the shape distribution consists of galaxies with various shapes.

\indent \citet{SanchezJanssen2019} found that the low-luminosity (-10 mag > M$_{g}$ > -15 mag) galaxy disks become thicker with decreasing total luminosity. According to that work, this result seems to hold regardless of the galaxy morphology, although absolute thickness of the disks of different morphological classes differ. Similar results were obtained by \citet{Kado-Fong2021}. These results are compatible with our findings that low-mass galaxies appear not to be thin disks, but we do not find an increasing trend in the mean $b/a$ distributions (Fig. \ref{fig:ba_in_mass_bins}), which would indicate a systematic thickening of the galaxies.

\indent \citet{Su2021} analyzed the FDSDC galaxies using multicomponent photometric decompositions and found that the fraction of galaxies with multiple components, such as a disk and a bar or a disk and a bulge, decreases toward lower masses. A similar finding for dwarf galaxies in the Virgo cluster was made by \citet{Janz2016}. Additionally, Su et al. showed that the fraction of galaxies described well with a single S\'ersic component also increases for the Spitzer Survey of Stellar Structure in Galaxies (S$^4$G, \citealp{Sheth2010}, \citealp{Salo2015}) toward lower masses. This disappearance of disk or multicomponent features happens around -18 mag > M$_{r'}$ > -16 mag, which is the same range where the correlation between the observed $b/a$ and $\bar{\mu}_{e,r}$ in our sample disappears. These findings may be an indication that the galaxies less luminous than that may have more dispersion-dominated thick "disks" that are not cool enough to be forming typical disk structures found in the larger disk galaxies. If we assume that this interpretation is true, it would have significant consequences for the kinematical analysis of the dwarf galaxies.

\subsection{Connection between size and other properties}

In Sect. 6.2 we showed that the SB of galaxies depends on projection effects and on its stellar populations. However, even after transforming the stellar light into stellar mass and correcting for the projection effects galaxies at a given mass have a large range of surface densities and thus different sizes. Taking into account the complexity of galaxy evolution, it is a rather unsurprising result that present day galaxies of a given mass may have a range of sizes. Anyways, at the same time, it is possible that some galaxies could be out of equilibrium and their sizes temporarily increased due to tidal interactions with other galaxies and the cluster potential.

\indent Using only the currently available data, it is not possible to know the individual 3D positions and velocities of the galaxies and calculate the exact tidal forces acting on them. The tidal forces are strongest in the center of the cluster, and thus the projected distance from the cluster center can be statistically used as a proxy for the tidal forces.  

\indent Tidal forces affect the dwarf galaxy structure and morphology, so we can try to identify whether we can detect such signs in the Fornax cluster dwarfs. Possible effects of tidal interactions are distortions in the shapes of the galaxies (i.e., increased elongation and possibly tidal tails and arms). Since tidal forces act in the radial direction, during and immediately after the tidal encounters, elongation of interacting galaxies tend to be aligned toward the source of the tidal force. This may manifest into the alignment of the galaxies in the cluster, so the tidally perturbed galaxies may be preferably elongated toward the cluster center. If the effects of the tidal forces are significant at the scale of the whole dwarf population, we should also detect more elongated galaxies in the inner parts of the cluster and find more tidally disturbed morphologies within those galaxies than those in the cluster outskirts.

\indent In Fig. \ref{fig:excess_size} we show the cumulative distributions of cluster-centric distance, cluster-centric alignment angle, axis ratio, and color for dwarfs. We divided the sample into three parts based on the excess size (R$_e$) of the galaxies with respect to mean size at given mass. The three groups considered are: galaxies that deviated by less than 1$\sigma$ from the distribution mean (blue lines), galaxies that deviate by more than 1$\sigma$ (green lines), and galaxies that deviate by more than 2 $\sigma$ (red lines). Assuming that the environment or tidal interactions do not play a role in the sizes of the galaxies, we should not find any differences between the samples. However, we do find differences in the properties of the normal galaxies and the galaxies with large R$_e$: in addition to the concentrated distribution of the large R$_e$ galaxies (K-S test for the assumption of similar distributions p=0.002) in the cluster, they seem to also have a slight tendency to have their major axis preferably aligned toward the cluster center (p=0.09 for more than 2 $\sigma$ deviators) and have larger fraction of galaxies with particularly small axis ratios (p=0.01). There is no significant difference between the colors of the different samples. 

\indent The radial alignment of dwarf galaxies in the Fornax cluster was also identified by \citet{Rong2019}. However, they did find the alignment signal for all the dwarf galaxies independent of their size. We do not find a significant signal for all dwarfs, but only a weak preferable alignment for the ones with large excess size.

\begin{figure*}
    \centering
    \includegraphics[width=16cm]{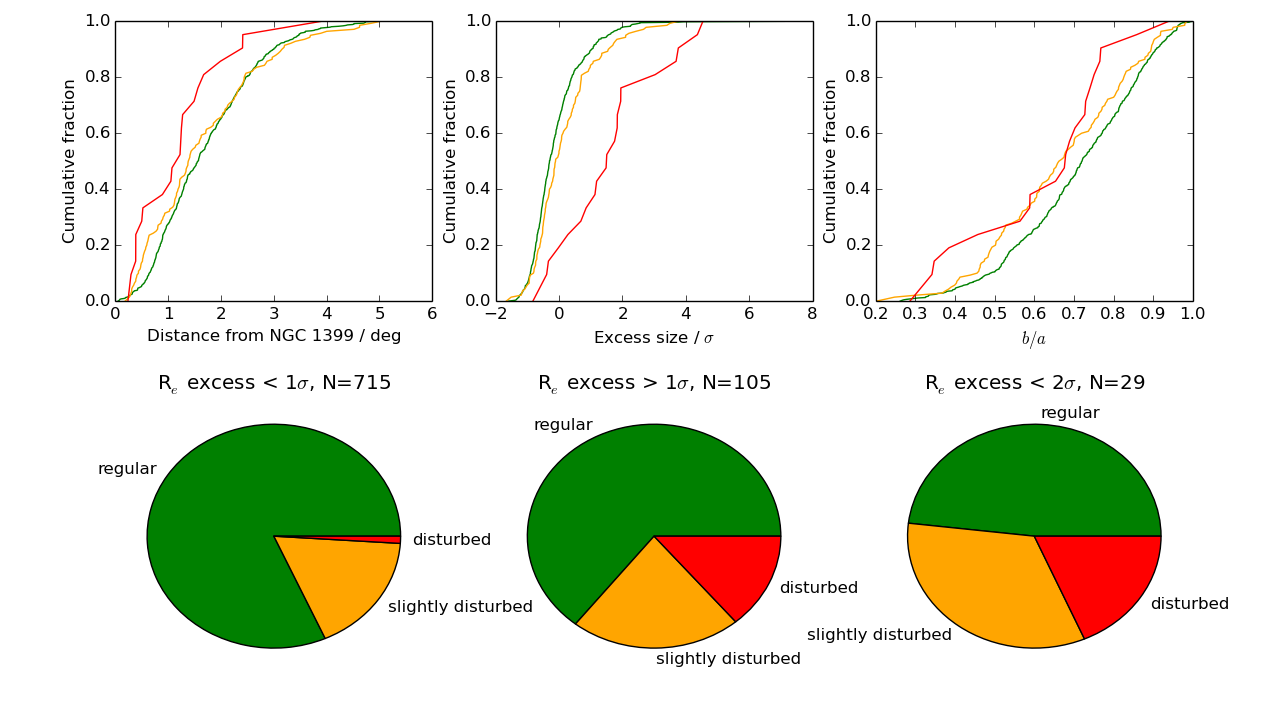}
    \caption{Upper panels: Cumulative distributions of cluster-centric distance (left panel), excess size (middle panel), and axis ratios (right panel) for the galaxies with different tidal morphologies. The colors of the curves in the panels correspond to the morphological classes named in the bottom panel pie charts and indicated with the same colors. Bottom panels: Pie charts showing the fractions of the tidal morphological classes among galaxies that deviate by less than 1$\sigma$ from the mass normalized R$_e$ distribution mean (left), galaxies that deviate by more than 1$\sigma$ (middle), and galaxies that deviate by more than 2 $\sigma$ (right). }
    \label{fig:tidal_class_params}
\end{figure*}

\begin{figure}
        \includegraphics[width=\columnwidth]{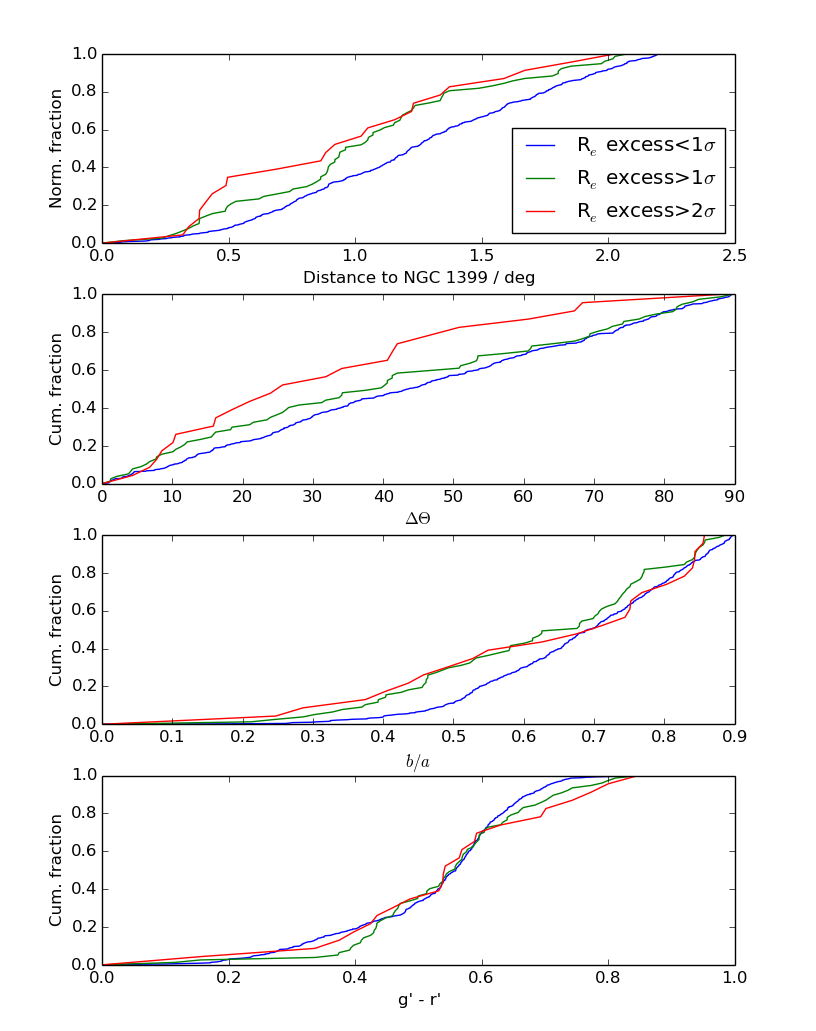}
        \caption{Cumulative distributions of cluster-centric distance (top panel), cluster-centric orientation (second panel), axis ratio (third panel), and color (bottom panel) for the Fornax cluster dwarfs. The different colors indicate the galaxies with different excess sizes: The blue lines show the galaxies with excess R$_e$ < $\langle \mathrm{R_e(M_*)} \rangle$ + 1$\sigma$, the green lines galaxies with R$_e$ > $\langle\mathrm{R_e(M_*)}\rangle$ + 1$\sigma$, and the red lines galaxies with R$_e$ > $\langle\mathrm{R_e(M_*)}\rangle$ + 2$\sigma$.}
        \label{fig:excess_size}
\end{figure}

\indent In addition to the previous analysis, we can also use the morphological classifications (Sect. 5) of tidal features in dwarfs to identify effects of the cluster's tidal forces. In Fig. \ref{fig:tidal_class_params}, we show the distributions of cluster-centric distances, excess size, and axis ratios for the different morphological types and show the appearance of the different tidal morphologies among the different excess size classes discussed before. We find that the cluster-centric distances decrease and excess sizes of the dwarfs increase from the galaxies with regular morphology to possibly disturbed and disturbed galaxies. The K-S test shows significant difference between the distributions of all the studied parameters for morphologies one (regular) and two (slightly disturbed, p=0.02, 0.02, 0.02 for distance, axis ratio, and excess size, respectively). The differences are also significant between classes one and three (clearly disturbed, p=0.07, 0.03, 1e-7 for distance, axis ratio, and excess size, respectively). The two disturbed classes only differ significantly by their excess size (p=2e-5).

\indent The results of the previous analyses are compatible with the idea that tidal forces drive dwarfs out of dynamical equilibrium in the inner parts of the cluster. Based on our calculations (\citealp{Venhola2019}), galaxies need to either travel closer than a few hundred parsec from the NGC 1399 in order to have the tidal radius within 2 R$_e$ from the center or spend a few gigayears in the cluster to experience similar stripping due to harassment. However, with the available information we cannot identify with certainty whether these tidal-forces whose effects we identify in galaxies are related to the cluster potential or interactions between individual galaxies.

\subsection{Formation of UDGs and LSB dwarfs}

In Sect. 6 we analyzed the properties of dwarf subpopulations defined using their mass, SB and surface density and size. Regardless of the used nomenclature (that we adopted here just for the sake of comparing the distributions), there seems to be no bimodal clustering in the galaxy properties that would give a physical justification to assign separate names "LSB-dwarf" or "UDG" (see, e.g., Figs. \ref{fig:dist_in_clust}, \ref{fig:sfd_correlation}, and \ref{fig:mag_mu_sample}). In general, the SD of dwarfs drops with decreasing total luminosity; in addition to that dependence, at any given mass there are deviations from that trend that approximately follow a normal distribution with a standard deviation of 0.5 dex in the surface densities. As seen in Fig. \ref{fig:sfd_correlation}, some of the UDGs deviate by less than 1 $\sigma$ from the mean $\bar{\mu}_{e,r'}$ of the dwarfs and are thus typical dwarfs. In general, the names LSB dwarf and UDG are misleading and should not be understood as separate types of dwarf galaxies, but rather as referring to a normal galaxy with low SB. In light of our data, any SB limit for those classes seems arbitrary.

\indent The formation mechanism of UDGs has been investigated in many theoretical studies during the last years (\citealp{Sales2019}, \citealp{Jiang2019}, \citealp{Tremmel2020}, \citealp{Amorisco2016}, \citealp{DiCintio2016}). Based on these works, UDGs are a subpopulation of dwarf galaxies that form naturally in isolation (e.g., \citealp{Amorisco2016}, \citealp{DiCintio2016}, \citealp{Sales2019}), but they are also produced in galaxy clusters via tidal forces and fading after quenching of their star-formation (\citealp{Sales2019}, \citealp{Jiang2019}, \citealp{Tremmel2020}).

\indent According to simulations, the UDGs formed via the last environmental effects are a $\approx 60\% $-majority, which leads to the theoretical prediction that UDGs in clusters are more centrally concentrated in the cluster than other dwarfs (\citealp{Sales2019}, \citealp{Jiang2019}). Our analysis of the distributions of UDGs and dwarfs in the Fornax cluster is consistent with this theoretical prediction. If we compare the fraction of UDGs of the total dwarf galaxy population at different cluster-centric radii, within D < 0.5R$_{vir}$ we get 14 UDGs out of 251 dwarfs = 5.6$\pm$1.4\%, within D < R$_{vir}$: 24/557 = 4.3$\pm$0.8\%, and D > R$_{vir}$ 8/221 = 3.6$\pm$1.2\%. However, the statistics are not strong enough to give support for the significance of the differences.

\indent The tidal forces in the cluster center do not affect only UDGs but the whole dwarf galaxy population. As discussed in Sect. 7.4, this phenomenon may also explain our finding that the RLSB dwarfs are more frequently disturbed, are distributed closer to the cluster center than the normal dwarfs, tend to have smaller apparent axis ratios than others, and that their major axes are preferably aligned toward the cluster center

\indent According to \cite{Jiang2019} and \cite{Sales2019}, the tidally enlarged UDGs fell into the cluster earlier than the majority of the dwarf population. This fact makes mean ages of their stellar populations older than the average. This is consistent with our observation that UDGs tend to have redder-than-average colors (see Fig. \ref{fig:sfd_correlation}). However, according to, for example, \cite{Sales2019}, cluster UDGs that are not tidally enlarged but are large LSB galaxies in equilibrium falling for the first time into the cluster should also comprise  40\% of the population. 

\indent \citet{Tremmel2020} showed that it is not only the tidal enlargement of the effective radii of galaxies, but also their fading stellar populations after the quenching of the star formation happens that makes the SB of galaxies fade in galaxy clusters. As a result from this physical process, Tremmel et al. find a correlation between the central SB and stellar age in the inner parts of dwarfs. We discussed this scenario in the Fornax cluster in our previous paper \citet{Venhola2019}, and showed that it is likely that all the galaxies with M$_*$ < 10$^8$ M$_*$ are quenched by RPS when they enter the inner parts of cluster for the first time. We also showed that the correlation between the $\bar{\mu}_{e,r'}$ and colors of the dwarfs are consistent with the fading scenario when studied in stellar mass bins. This correlation has been afterward identified also in the Virgo and Centaurus cluster dwarfs by \citet{Janz2021}. 

\begin{figure}
        \includegraphics[width=\columnwidth]{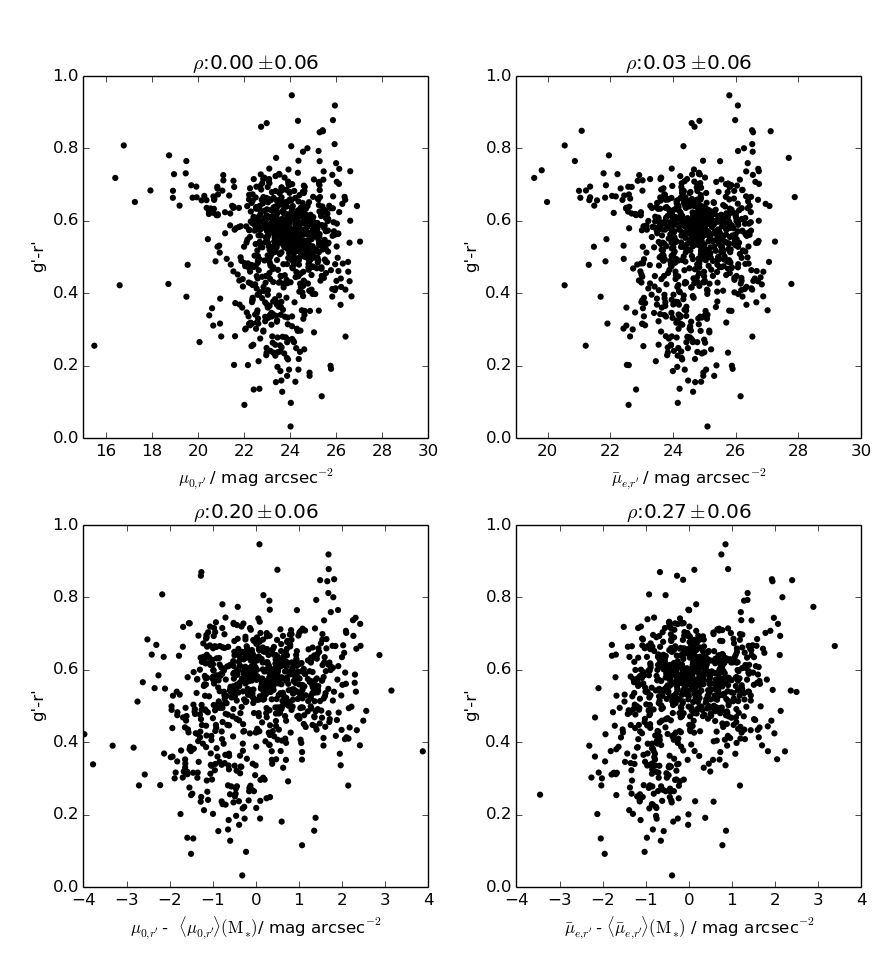}
        \caption{Upper panels: Galaxy g'-r' colors as a function of their central (left) and mean effective (right) SBs. Lower panels: Same measures, but the SBs are compared with the mean value at a given mass. Spearman rank correlation coefficients with their uncertainty are listed above the panels.}
        \label{fig:color_sfb_all}
\end{figure}

\indent To test whether the correlation found in the simulations of Tremmel et al. still holds in the Fornax cluster after including the dwarf galaxies identified in this work to the analysis, we plot the SB versus color relation in the upper panels of Fig \ref{fig:color_sfb_all}. We find that there is no significant correlation between the absolute SB and colors of the galaxies. However, there exists the luminosity-SB relation for dwarfs, which needs to be taken in account. In the lower panels of Fig. \ref{fig:color_sfb_all} we compare the SBs with the mean SB at given mass and find the significant correlation between the color and SB.

\begin{figure}
        \includegraphics[width=\columnwidth]{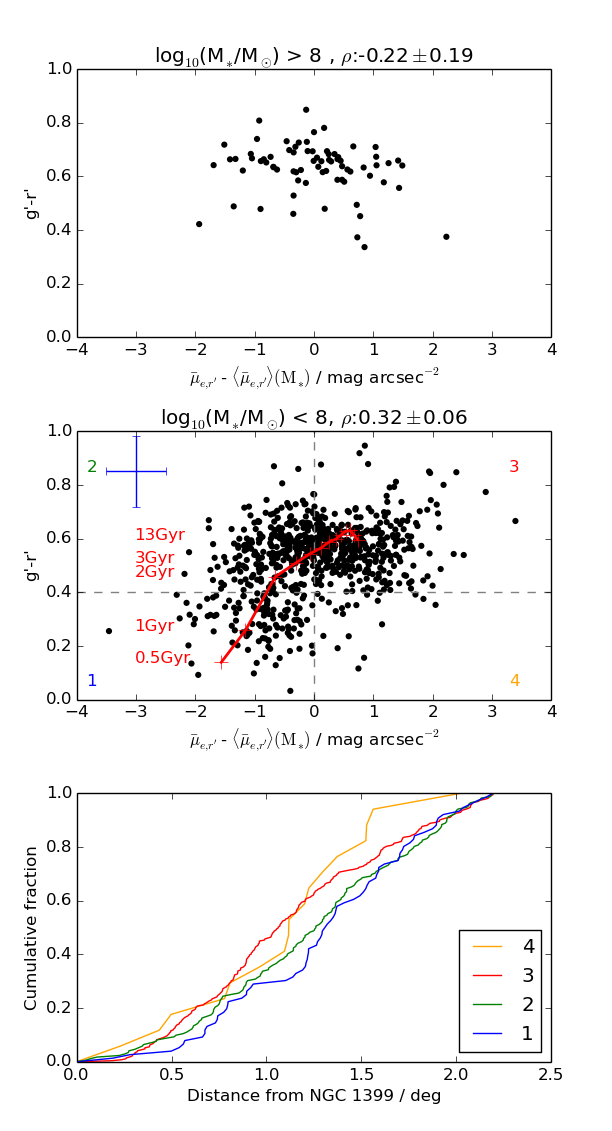}
        \caption{Two upper panels: Galaxy g'-r' colors as a function of their excess mean effective SB. Top and middle panels:  Dwarf galaxies with log10(M$_*$/M$_{\odot}$) > 8 and log10(M$_*$/M$_{\odot}$) < 8, respectively, shown separately. Spearman rank correlation coefficients with their uncertainty are listed above the panels. In the middle panel, we also present a stellar population evolution model that shows the color and SB evolution of an aging stellar population (see text for details). The stellar population ages corresponding to different colors are shown in red on the left. The blue cross in the middle panel shows the average uncertainties in the parameters. Bottom panel: Cluster-centric distributions of the dwarfs located in the different quadrants of the middle panel. }
        \label{fig:color_sfb_dwarfs}
\end{figure}

\indent The dwarfs with log$_{10}$(M$_*$/M$_{\odot}$) > 8 follow a different relation than the less massive ones (upper panel of Fig. \ref{fig:color_sfb_dwarfs}), so we exclude them from the further comparisons. In the mid panel of Fig. \ref{fig:color_sfb_dwarfs}, we compare the remaining dwarfs with a single stellar population evolution track using the models of \citet{Vazdekis2010}. We use a Kroupa initial mass function (IMF), and show the model for stellar population with metallicity of log$_{10}$(Z/Z$_{\odot}$)=$-1.0$. As the x axis shows the mass-dependent excess SB, the location of the stellar evolution track on the x axis is arbitrary. As shown by Venhola et al. (2019) in the Fornax cluster and \citet{Janz2021} for the Virgo cluster dwarfs, the stellar population evolution can fit the SB and color distribution of most of the galaxies. This is also the case after including the LSB extension. The scatter in the color axis is consistent with the typical 0.1-0.2 mag uncertainties of the colors. 

\indent The degeneracy between age and metallicity in the optical colors causes some difficulty in straightforwardly interpreting the colors. However, the change in metallicity from log$_{10}$(Z/Z$_{\odot}$)=0 to -1.5, which covers the expected range for dwarfs well, only has a 0.3 dex change in the colors, which is on the order of measurement uncertainties. The change in metallicity does not have a significant effect on the SB evolution, so even if there is a range of metallicities among the dwarfs, a similar trend in the color-SB plane is expected (\citealp{Janz2021}).

\indent If the SB and color distribution of dwarfs manifests  the rapid quenching of star formation in the cluster environment and subsequent aging of the stellar populations, we should find a trend in the cluster-centric distributions of the dwarfs with different quenching times. In order to check this, we divided the excess SB-color relation into four quadrants (mid panel of Fig. \ref{fig:color_sfb_dwarfs}) and plotted the cluster-centric distributions in the bottom panel of Fig. \ref{fig:color_sfb_dwarfs}. If the locations of galaxies in the excess SB-color relation are caused by their evolution along the evolutionary curve of the stellar population, we should observe galaxies in the quadrant one to be the outermost population from the cluster center, galaxies in the second in the middle, and galaxies in the quadrant three as the innermost population. This is the case as it is shown in the bottom panel of Fig. \ref{fig:color_sfb_dwarfs}. 

\indent The galaxies in the fourth quadrant of Fig. \ref{fig:color_sfb_dwarfs} are hard to explain with the model relying solely on the evolution of stellar populations: they are blue galaxies with particularly low SB. However, as discussed in Sect. 7.4, tidal forces likely cause disturbances in some galaxies making their SB fainter than typical. If this assumption is true we should find those galaxies distributed in the inner parts of the cluster. As seen in the bottom panel of Fig. \ref{fig:color_sfb_dwarfs}, quadrant four galaxies are the innermost population together with the quadrant three galaxies. The differences in the distributions are significant between the galaxies in quadrants one and three (K-S test p=0.0003) and two and three (P=0.02)

\begin{figure}
        \includegraphics[width=\columnwidth]{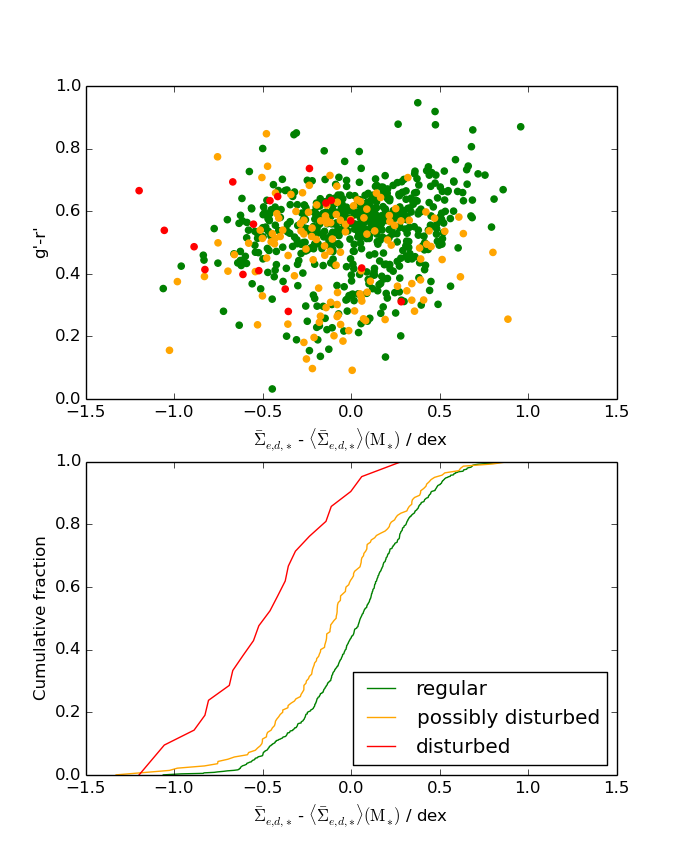}
    \caption{Upper panel: Galaxy g'-r' colors as a function of excess SD. The different colors correspond to galaxies with different amounts of disturbed morphology. Lower panel: Cumulative distributions of the excess SD of the different morphological classes of the upper panel.}
                \label{fig:color_sfb_dwarfs_tidal}
\end{figure}

\indent Since both the age fading of stellar populations and tidal interactions affect the SB of a galaxy, but only the latter alters the surface mass density of the galaxy, it makes sense to recreate the middle panel of Fig. \ref{fig:color_sfb_dwarfs} using surface mass density instead of brightness. In the upper panel of Fig. \ref{fig:color_sfb_dwarfs_tidal}, we show the excess SD-color relation for the Fornax cluster dwarfs. Now, the galaxy's location on the  color axis depends on the quenching time and the excess SD can be increased by the tidal forces. We acknowledge that taking into account those two processes is not enough to uniquely define a galaxy's location on the given axis, but if the tidal interactions have significance, we should observe the tidally disturbed galaxies to have lower excess SD compared to galaxies with similar colors. In the upper panel of Fig. \ref{fig:color_sfb_dwarfs}, we show galaxies with different amounts of morphological disturbance with different colors and show their cumulative distributions in the lower panel. We find that galaxies with disturbed morphology indeed tend to have lower SD than the typical galaxies at a given color. The differences between all the morphological classes are significant (K-S test p < 0.002).

\indent These observational results are well in agreement with the theoretical frameworks that \citet{Sales2019}, \citet{Jiang2019}, and \citet{Tremmel2020} have described for the formation of the UDGs. However, here we show that the stellar population aging and tidal interactions can be used to understand the formation of RLSB/RLSD dwarfs in the Fornax cluster in the whole dwarf population with log10(M$_*$/M$_{\odot}$) < 8. 

\section{Summary and conclusions}

In this work we have quantified the detection efficiency and selection effects of the MTO using simulated images and the FDS data. The detection efficiency of MTO by far exceeds that of SE and thus justifies its use for LSB galaxy detection. We found that the MTO is able to increase the detection completeness in the FDS data by reaching 0.5-1 mag arcsec$^{-2}$ fainter SBs; additionally, by detecting more accurate sizes for the galaxies, it decreases the amount of galaxies that get excluded from the sample by the size cuts. We then combined MTO with an automated galaxy fitting pipeline that applies GALFIT to fit galaxies with S\'ersic profiles. As a result, we obtained an LSB galaxy sample that extends the previous FDS dwarf galaxy catalog of the Fornax cluster by V18. Using the new catalog, we studied the low-luminosity end of the LF in the Fornax cluster and analyzed the structural trends of the dwarfs in the cluster. Our main results are the following:

\begin{itemize}

\item We identified 265 new LSB dwarf candidates in the Fornax cluster that were not included in the FDSDC. Extending the FDSDC with the galaxies of this work increases its total number of galaxies to 821. This galaxy sample is more than 75\% complete down to M$_{r'}$ = -12 mag and $\bar{\mu}_{e,r'}$ = 26 mag arcsec$^{-2}$ and reaches its 50\% completeness limits at M$_{r'}$ = -10 mag and $\bar{\mu}_{e,r'}$ = 26.5 mag arcsec$^{-2} (Fig. 1)$. The limits are 0.5-1 mag deeper than for the FDSDC. The faintest galaxies of the sample have stellar masses of M$_*$ = 5$\times$10$^4$ M$_{\odot}$.
\vskip 0.3cm

\item The faint-end LF in the Fornax cluster has a slope of $\alpha$=-1.38 $\pm$ 0.02 down to the completeness limit of our sample, M$_{r'}$ = -12 mag (Fig. 10). Our sample is size limited and thus excludes extremely compact galaxies, such as Ultra-Compact Dwarfs  (UCD) and dwarfs with $a$ < 200 pc. 
\vskip 0.3cm

\item When galaxies are classified based on their SB or density, the galaxies that are sparser than average are more centrally clustered in the Fornax cluster than the dense galaxies (Fig. 13). Ultra-diffuse galaxies, being the innermost population of the cluster dwarfs in their mass range (as long as UCDs are not considered), follow the same trend.
\vskip 0.3cm

\item Our observation that the excess size of the galaxies with respect to the average galaxy at a given mass increases toward the center of the cluster (Fig. 17) is consistent with the theoretical prediction that the tidal forces of the cluster potential input energy to the galaxies and thus increase their sizes. Galaxies with large excess sizes tend to also be more elongated than normal galaxies, and their major axes are preferably aligned toward the cluster center (Fig. 17). We also observe that the number of galaxies with disturbed morphologies increases toward the central parts of the cluster. These observations give further support to the importance of tidal forces in the cluster.
\vskip 0.3cm

\item We find no evidence for variations in the low-luminosity end of the LF between the Fornax cluster and the different galaxy environments. This result holds for galaxies with -18.5 mag < M$_V$ < -12.5 mag (Table 2 and Fig. 15).  
\vskip 0.3cm 

\item The low-mass slope of the galaxy stellar mass function in the Fornax cluster is steeper than in the cosmological simulation IllustrisTNG (Fig. 16). As these discrepancies become more severe toward the resolution limits of IllustrisTNG, it is likely that the differences are at least partly numerical.
\vskip 0.3cm

\item We find that there is a correlation between the apparent SB and axis ratio of dwarf galaxies with log$_{10}$(M$_*$/M$_{\odot}$) > 7 (Fig. 11), which indicates that they are disk-like. This correlation disappears for the less massive galaxies. This indicates that their shapes are not well approximated by disks.
\vskip 0.3cm

\item Based on the analysis of the dependence between the location, SB, colors, and morphology of dwarfs in the Fornax cluster (Fig. 19), we find that their properties can be explained with the theoretical models of the evolution of dwarfs within a cluster environment (\citealp{Sales2019}, \citealp{Jiang2019}, \citealp{Tremmel2020}). The effects of the two dominant mechanisms, tidal interactions and the aging of the stellar populations, can both be identified in the photometric properties of the dwarf galaxies.

\end{itemize}

{\bf Acknowledgements:} A.V. would like to thank the Emil Aaltonen Foundation for the financial support during the writing of this paper. R.F.P., E.L., H.S., E.I., M.H.F.W. , C.H. and J.J. acknowledge financial support from the European Union’s Horizon 2020 research and innovation programme under the Marie Skłodowska-Curie grant agreement No. 721463 to the SUNDIAL ITN network. H.S., E.L., and A.V. are also supported by the Academy of Finland grant n:o 297738.

\bibliographystyle{aa}

\bibliography{dwarf_paper_det}

\begin{appendix}

\section{Details of catalog production}

\subsection{Pixel value distribution moments}

MTO and SE both measure semimajor and semiminor axis lengths for the objects. For calculating these parameters, second-order pixel value distribution moments are used. If we mark x and y coordinates and intensity of a pixel i as $x_i, y_i$ and $I_i$, respectively, we can calculate the second-order moments as follows
\begin{equation}
\begin{matrix}
\overline{x^2} = \frac{\sum_{i} I_i x_i^2 }{\sum_i I_i} - \overline{x}^2, \\ 
\overline{y^2} = \frac{\sum_{i} I_i y_i^2 }{\sum_i I_i} - \overline{y}^2,  \\
\overline{xy} = \frac{\sum_{i} I_i x_i y_i }{\sum_i I_i} - \overline{x} \, \overline{y}.
\end{matrix} 
\end{equation}
The second-order moments measure the spread along the x and y axes, and they are transformed to semiminor axis $b$ and semimajor axis $a$ as
\begin{equation}
\begin{matrix}
a^2  =\frac{\overline{x^2} + \overline{y^2}}{2} + \sqrt{\left(\frac{\overline{x^2} - \overline{y^2}}{2}\right)^2+\overline{xy}^2}, \\ 
b^2  =\frac{\overline{x^2} + \overline{y^2}}{2} - \sqrt{\left(\frac{\overline{x^2} - \overline{y^2}}{2}\right)^2+\overline{xy}^2},
\end{matrix} 
\end{equation}
and the position angle $\theta,$
\begin{equation}
\tan(2\theta) = 2 \frac{\overline{xy}}{\overline{x^2}-\overline{y^2}}.
\end{equation}

\begin{figure}
        \includegraphics[width=\columnwidth]{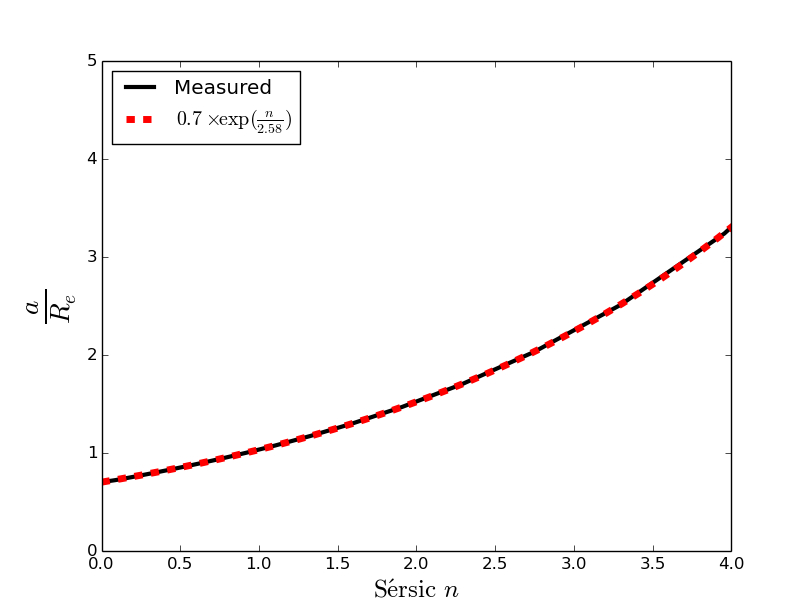}
    \caption{Ratio of $a$ and R$_e$ as a function of S\'ersic index for a pure S\'ersic profile.The black curve shows the values calculated from moments, and the dashed red line shows the analytical approximation.}
    \label{fig:a_reff}
\end{figure}

\indent In general, the major-axis $a$ calculated from the pixel value distribution moments is not straightforwardly comparable with R$_e$. Their ratio is dependent on the concentration (i.e., the S\'ersic index) of the profile. In order to give the reader insight how these two are related, we calculated the ratio between $a$ and R$_e$ for a pure S\'ersic profile and show their ratio as a function of $n$ in Fig. \ref{fig:a_reff}. From the figure we see that the ratio of the two size measures is close to unity for the S\'ersic $n$ values typical for dwarfs (0.5 < $n$ < 2). We also find that the relation can be well approximated by a fitting formula:
\begin{equation}
    \frac{a}{\mathrm{R}_e} \, = \, 0.7 \times \exp \left(\frac{n}{2.38} \right)
.\end{equation}

\subsection{SE configuration parameters}

Configuration parameters of SE are described in detail in the SE guide that is available at \url{https://www.astromatic.net/pubsvn/software/sextractor/trunk/doc/sextractor.pdf}. The most important parameters for the object detection are the size of the background grid, detection threshold and the minimum area of the detections. One can also apply different convolution filters for images, but in order to keep the comparison simple, we used a simple 3 pix $\times$ 3 pix Gaussian filter with full width half maximum (FWHM)  2 pix, which is also used by MTO. In order to decide which parameter values we should adopt for the detection tests we ran a test for a set of parameters. We selected a few reasonable values for each parameter and ran SE using all combination of those parameters for one of the mock images described in Sect. 2.2. We used values of 500, 1000, and 10000 pixels for the background grid size, 0.5 and 1$\sigma$ for the detection threshold, and 10, 50, and 100 pixels for the minimum detection area. We then selected the set of parameters that provided the best combination of completeness and purity of the detections. The optimal performance was obtained using a background size of 500 pixels, a detection threshold of 1$\sigma$, and a minimum detection area of 10 pixels. We then used the following set of configuration parameters when running SE:

\begin{tiny}
\begin{verbatim}
DETECT_MINAREA   10,          # minimum number of pixels above 
                              # threshold
DETECT_THRESH    1,           # <sigmas> 
ANALYSIS_THRESH  1,           # <sigmas> 
FILTER           Y,           # apply filter for detection 
DEBLEND_NTHRESH  64,          # Number of deblending 
                              # sub-thresholds
DEBLEND_MINCONT  0.005,       # Minimum contrast for deblending
CLEAN            Y,           # Clean spurious detections? 
CLEAN_PARAM      1.0 ,        # Cleaning efficiency
MASK_TYPE        CORRECT,     # type of detection MASKing :
                              # NONE, BLANK or CORRECT 
MAG_ZEROPOINT    0.0,         # magnitude zero-point
PIXEL_SCALE      0.2,         # size of pixel in arcsec 
SEEING_FWHM      1.2,         # stellar FWHM in arcsec
STARNNW_NAME     default.nnw, # Neural-Network_Weight 
                              # table filename
BACK_SIZE        500,         # Background mesh <size> 
BACK_FILTERSIZE  3,           # Background filter <size>
BACKPHOTO_TYPE   GLOBAL,      # can be GLOBAL or LOCAL 
MEMORY_OBJSTACK  3000,        # number of objects in stack
MEMORY_PIXSTACK  3000000,     # number of pixels in stack
MEMORY_BUFSIZE   1024,        # number of lines in buffer 
\end{verbatim}
\end{tiny}
\subsection{Detection completeness }
\begin{figure*}
        \includegraphics[width=17cm]{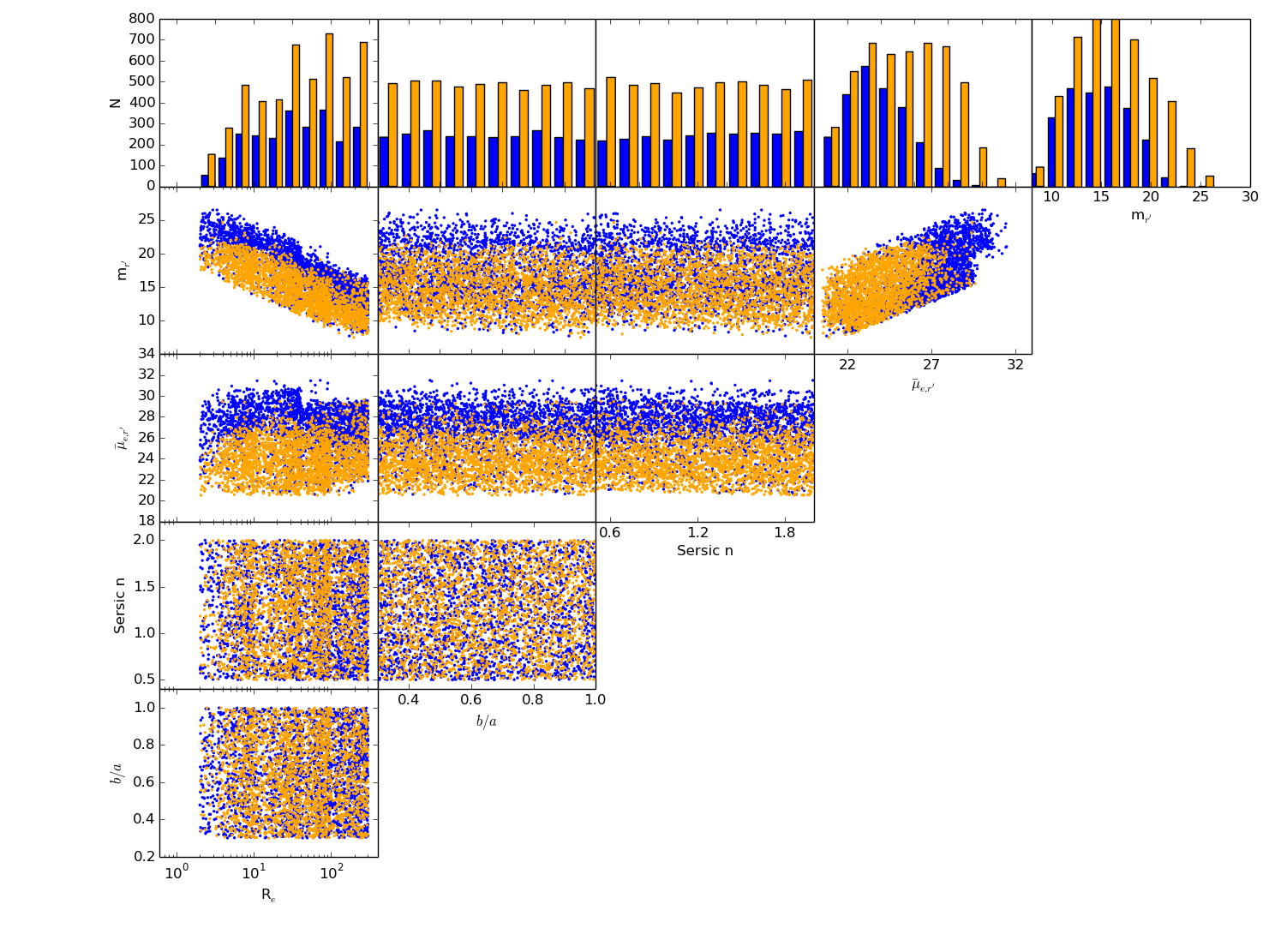}
    \caption{Detections ({\it orange points}) and non-detections ({\it blue points}) of the mock galaxies using MTO. The matrix plots show the distribution of galaxies in the different parameter subspaces. The upper histograms show the distributions of detected (orange) and all (blue) galaxies for the given parameters.}
        \label{fig:full_Det_an}
\end{figure*}

In Fig. \ref{fig:full_Det_an} we show how the detections and the input galaxy properties are distributed in the different parameter spaces. 

\subsection{Examples of detection artifacts}

MTO forms the max-tree by identifying sets of connected pixels with different thresholds and then labels the branches according to the rules given in Sect. 3. As MTO does not assume any profile for the detected objects, some parts of the diffuse light are often associated with a wrong object. This is especially the case with interacting galaxies and densely grouped galaxies. Their light profiles are connected at low levels and get associated with one of the galaxies, thus making the galaxy's mean SB low. 

\indent Another type of low SB features that gets identified by MTO but is considered as an artefact in our LSB dwarf catalog are stellar reflection halos. As shown by V18, the inner reflections of the VST cause the PSF to have reflection rings around bright objects. The SB of those rings is $\approx$ 10 mag arcsec$^{-2}$ fainter than the peak of the PSF and have angular size on the order of 5 arcmin. The SB and size of these features are similar to LSB dwarfs, but they are easy to separate from each other based on their morphology and the fact that reflection rings appear near bright stars. Most of the false positives caused by the reflection rings can be identified using stellar catalogs of the Fornax area.

\begin{figure}
        \includegraphics[width=\columnwidth]{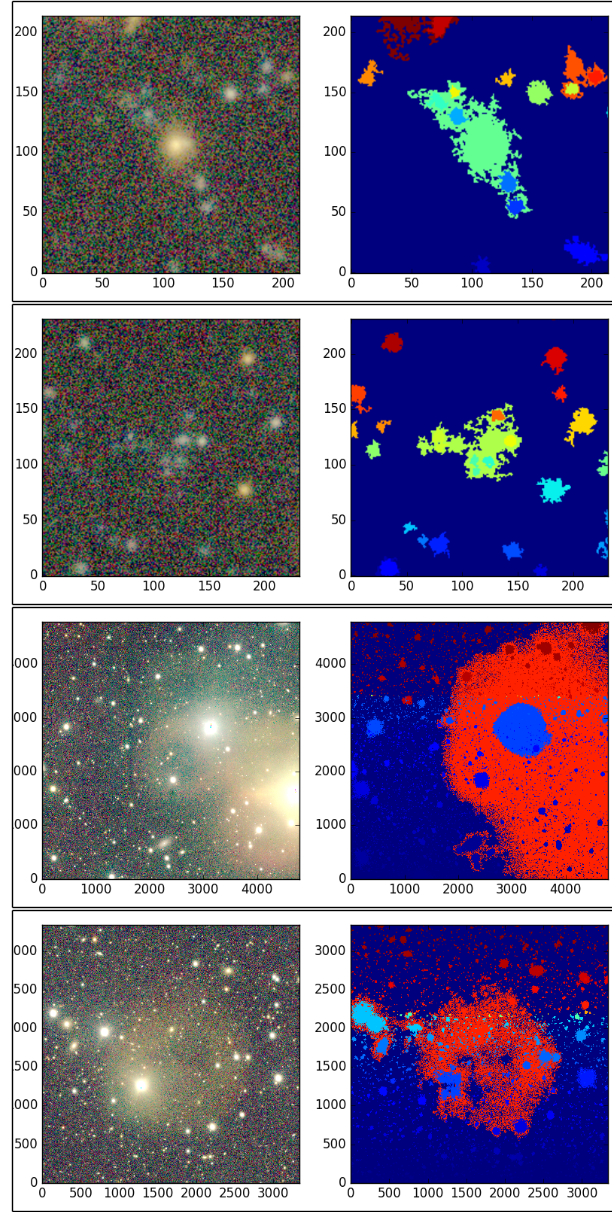}
    \caption{Examples of sources that appear as LSB galaxies in the MTO detection but have different physical natures. The left panels show color images generated from the FDS data, and the right panels show the segmentation maps, where different object labels are shown with different colors. The two upper examples show groups of galaxies that are connected on low SB levels and where the common LSB envelope is associated with one of the galaxies, making its mean SB low. The two lower panels show reflection rings around bright stars. Those objects have similar SBs as LSB dwarfs.}
    \label{fig:examples_of_difficult}
\end{figure}

\subsection{Identification of nuclei}

In order to automatically decide whether a galaxy needs to be fitted with a S\'ersic or S\'ersic+nucleus, we compare the Gini coefficient of central part of the galaxy with the Gini coefficient of the PSF. The calculation is done within the innermost two arcsec from the center of the galaxy and the PSF. The Gini coefficient describes the distribution of flux within the selected pixels and is defined as follows:
\begin{equation}
    G\, =\, \frac{\Sigma_{i=1}^{n}\Sigma_{j=1}^{n} |x_i-x_j|}{2n^2\bar{x}} ,
\end{equation}
where $x_i$ and $x_j$ correspond to the fluxes within the pixels $i$ and $j$, $n$ is the number of pixels, and $\bar{x}$ is the average flux within the pixels. Value of $G$=1 means that all the flux is in the brightest pixel and $G$=0 means that all the pixels have equal amount of flux.

\indent As $G$ describes how large fraction of the flux is in the brightest pixels, a point source has the maximum $G$ that any object can have. In the case of astronomical images, the signal is convolved with the PSF, so the fraction $G_{obj}$/$G_{PSF}$ tells how similar an object is to a point source. If we consider a typical dwarf galaxy in the Fornax cluster, its light profile is usually well described by a S\'ersic profile with 0.5 < $n$ < 1.5, and it may or may not have a point-like nucleus in the center. Clearly, if there is a point source that has high contrast with respect to the main body of the galaxy, the $G_{obj}$/$G_{PSF}$ will be close to one in the core. However, the nucleus can also be faint with respect to the S\'ersic component of the profile, or the profile may have high $n$, which makes it difficult to identify the nucleus. In order to test what would be a practical $G_{obj}$/$G_{PSF}$ threshold value for indicating likely presence of nucleus, we generated a set of simulated galaxies with different SB, S\'ersic $n$ and nucleus fraction. We then calculated $G_{obj}$/$G_{PSF}$ for all the galaxies. 

\begin{figure}
        \includegraphics[width=\columnwidth]{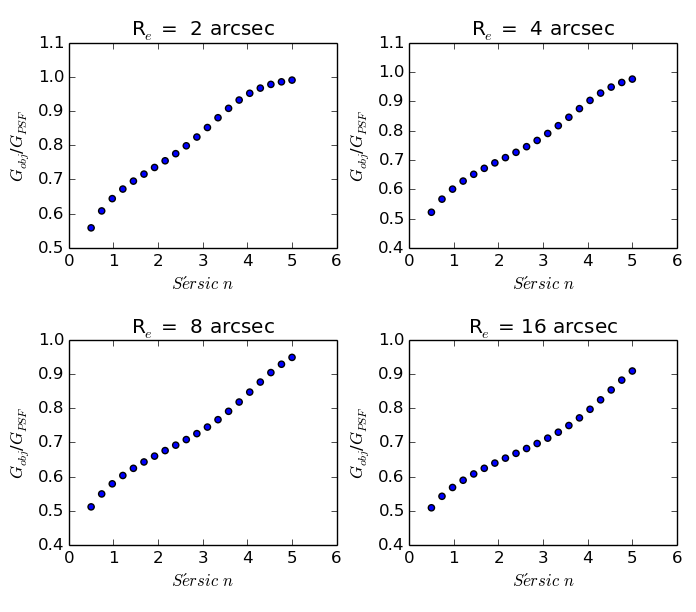}
    \caption{ $G_{obj}$/$G_{PSF}$ ratio as a function of S\'ersic index for non-nucleated galaxies. The different panels correspond to galaxies with different effective radii. }
        \label{fig:gini_reff_n}
\end{figure}

\indent First we studied how $G_{obj}$/$G_{PSF}$ behaves as a function of S\'ersic $n$  and R$_e$of the galaxy. We simulated noiseless galaxies with different $n$ and R$_e$ and show the results in Fig. \ref{fig:gini_reff_n}. Our results show that $G_{obj}$/$G_{PSF}$ is close to 0.5 -- 0.6 for an exponential profile and then grows close to one when $n$ grows. The size of the galaxy does not effect the value much as long as the galaxy is resolved. 

\begin{figure*}
        \includegraphics[width=17cm]{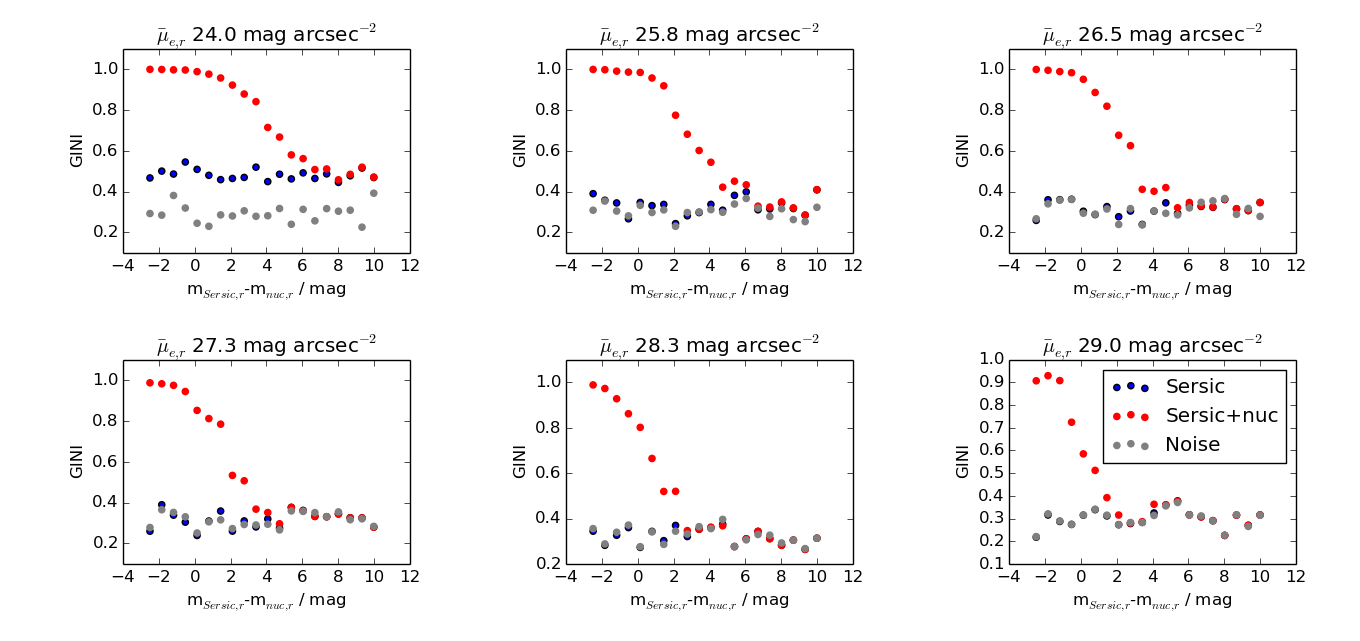}
    \caption{$G_{obj}$/$G_{PSF}$ calculated for an exponential profile (red symbols), a nucleated exponential profile (blue symbols), and noise (gray symbols). The values on the x axis indicate the difference between the total magnitude of the simulated nucleus and that of the galaxy, and thus those values are only relevant for the red symbols. For other symbols, the different locations on the x axis correspond to different realizations of the same galaxy. The panels show the $G_{obj}$/$G_{PSF}$ values for galaxies with different SBs.}
        \label{fig:gini_comp}
\end{figure*}

\indent In Fig. \ref{fig:gini_comp} we show the $G_{obj}$/$G_{PSF}$ ratio for nucleated and non-nucleated exponential galaxies with different SBs. In this simulation, we also add Gaussian noise to the images that corresponds to the FDS r'-band noise. In the case of the non-nucleated galaxy with the highest SB, the galaxy light dominates over the background noise and the $G_{obj}$/$G_{PSF}$ is close to 0.5. With decreasing SB, the value goes down and eventually reaches $G_{obj}$/$G_{PSF}$=0.3, which is identical to the value of noise. In the case of the nucleated galaxy, the $G_{obj}$/$G_{PSF}$ value is close to 1 when the fraction of light in the nucleus is large and reaches the same value as the non-nucleated galaxy when the fraction of light in the nucleus decreases. The nucleus fraction at which the nucleated and non-nucleated galaxies become indistinguishable is dependent on the SB of the galaxy. In the case of the brightest dwarfs we can detect a nucleus that has 1/100th of the luminosity of the S\'ersic profile, but in the case of the faintest dwarfs only a nucleus with luminosity larger than 1/10th of the S\'ersic profile can be identified using the $G_{obj}$/$G_{PSF}$ value.

\indent Based on these tests, we notice that $G_{obj}$/$G_{PSF}$ $\approx$ 0.3 for noise and LSB exponential profiles. The $G_{obj}$/$G_{PSF}$ ratio grows as S\'ersic $n$ grows, but since most of the dwarfs have 0.5 < $n$ < 1.5, the typical maximum value of $G_{obj}$/$G_{PSF}$ for a dwarf without a nucleus is around 0.5. However, for the non-nucleated dwarfs, the $G_{obj}$/$G_{PSF}$ ratio quickly drops below 0.4 with decreasing SB but remains above that value for nucleated dwarfs (as long as the nucleus can be identified from the data). Thus, we selected $G_{obj}$/$G_{PSF}$ = 0.4 as the threshold to include a nucleus into the GALFIT model.

\section{Catalog of LSB dwarfs}
Photometric parameters of the likely Fornax galaxies that were not identified by V18 are presented in an electronic catalog that is published with this paper. In Table \ref{tab:cat_example}, we present a small sample of the catalog to demonstrate its contents. 

\onecolumn
\begin{landscape}
\begin{table*}
\caption{ Example page of our dwarf galaxy catalog that includes likely cluster galaxies. The columns from left to right correspond to: target, right ascension (R.A.) and declination (Dec.) in ICRS coordinates, axis ratio ($b/a$), position angle measured from north to east ($\theta$), apparent r'-band total magnitude (m$_{r'}$), r'-band effective radius in arcseconds (R$_{e}$), S\'ersic index ($n$), aperture magnitudes within the effective radius in the u', g', r', and i' filters, the concentration index ($C$), and the residual flux fraction ($RFF$). The values after the $\pm$-signs correspond to the 1$\sigma$ uncertainties in the parameters. The last column tells the tidal morphological type of the galaxy according to our classifications in Sect. 5: ``1'' corresponds to a regular galaxy, ``2'' to a slightly disturbed galaxy, ``3'' to a clearly disturbed galaxy, and ``4'' to an unclear case, and asterisk at the end of the morphological classification indicates that the object has a nucleus. {\it Note:} The full version of this table is only available in electronic form at the CDS via anonymous ftp to cdsarc.u-strasbg.fr (130.79.128.5) or via \url{http://cdsweb.u-strasbg.fr/cgi-bin/qcat?J/A+A/}}
\label{tab:cat_example}
\centering
\resizebox{24cm}{!}{%
\begin{tabular}{lcccccccccccccc}
\hline\hline
Target & R.A. & Dec. & $b/a$ & $\theta$ & m$_{r'}$ & R$_{e}$ & $n$ & u' & g' & r' & i' & $C$ & $RFF$ & Morphology  \\
\hline

FDSLSB1 & 55.6888 & -35.4720 & 0.80 $\pm$ 0.06 & 10.6 $\pm$ 3.6 & 21.7 $\pm$ 0.2 & 2.0 $\pm$ 0.4 & 0.3 $\pm$ 0.4 & 23.43 $\pm$ 0.15 & 22.96 $\pm$ 0.13 & 22.78 $\pm$ 0.11 & 22.35 $\pm$ 0.07  & 1.9 & -0.00 & 2 \\ 
FDSLSB2 & 55.3600 & -35.3797 & 0.84 $\pm$ 0.06 & 52.5 $\pm$ 3.8 & 21.2 $\pm$ 0.3 & 2.7 $\pm$ 0.6 & 0.5 $\pm$ 0.4 & 24.52 $\pm$ 0.71 & 22.51 $\pm$ 0.14 & 21.90 $\pm$ 0.09 & 21.63 $\pm$ 0.07  & 2.9 & -0.02 & 1 \\ 
FDSLSB3 & 56.0367 & -35.1445 & 0.83 $\pm$ 0.03 & 32.8 $\pm$ 2.3 & 20.6 $\pm$ 0.2 & 2.0 $\pm$ 0.3 & 0.9 $\pm$ 0.3 & 23.62 $\pm$ 0.17 & 22.34 $\pm$ 0.08 & 21.69 $\pm$ 0.06 & 21.40 $\pm$ 0.05  & 2.7 & -0.00 & 1 \\ 
FDSLSB4 & 55.5467 & -35.4432 & 0.62 $\pm$ 0.09 & 61.5 $\pm$ 5.0 & 18.8 $\pm$ 0.3 & 11.2 $\pm$ 3.1 & 0.8 $\pm$ 0.4 & 21.43 $\pm$ 0.51 & 20.25 $\pm$ 0.18 & 19.72 $\pm$ 0.12 & 19.35 $\pm$ 0.09  & 2.2 & 0.00 & 1 \\ 
FDSLSB5 & 55.6311 & -35.0248 & 0.92 $\pm$ 0.09 & 24.7 $\pm$ 4.8 & 21.3 $\pm$ 0.3 & 3.3 $\pm$ 0.9 & 0.8 $\pm$ 0.4 & 23.78 $\pm$ 0.58 & 22.78 $\pm$ 0.27 & 22.19 $\pm$ 0.16 & 22.07 $\pm$ 0.13  & 2.5 & -0.05 & 1 \\ 
FDSLSB6 & 55.3456 & -35.8966 & 0.40 $\pm$ 0.11 & 10.7 $\pm$ 6.1 & 20.8 $\pm$ 0.3 & 5.5 $\pm$ 1.7 & 1.4 $\pm$ 0.4 & 23.62 $\pm$ 0.60 & 22.28 $\pm$ 0.20 & 21.73 $\pm$ 0.13 & 21.81 $\pm$ 0.12  & 2.6 & 0.01 & 1 \\ 
FDSLSB7 & 55.6970 & -35.3898 & 0.66 $\pm$ 0.10 & -39.7 $\pm$ 5.5 & 21.8 $\pm$ 0.3 & 3.1 $\pm$ 0.9 & 0.3 $\pm$ 0.4 & 24.27 $\pm$ 0.59 & 23.21 $\pm$ 0.26 & 22.68 $\pm$ 0.17 & 22.49 $\pm$ 0.12  & 2.4 & 0.00 & 1 \\ 
FDSLSB8 & 56.2119 & -35.7884 & 0.43 $\pm$ 0.14 & 88.4 $\pm$ 7.0 & 21.2 $\pm$ 0.4 & 5.4 $\pm$ 1.9 & 1.6 $\pm$ 0.4 & 23.91 $\pm$ 0.79 & 22.83 $\pm$ 0.33 & 22.40 $\pm$ 0.23 & 21.57 $\pm$ 0.10  & 2.9 & -0.04 & 1 \\ 
FDSLSB9 & 56.3158 & -35.2712 & 0.56 $\pm$ 0.06 & 9.5 $\pm$ 3.3 & 21.2 $\pm$ 0.2 & 2.4 $\pm$ 0.5 & 1.1 $\pm$ 0.4 & 23.68 $\pm$ 0.18 & 22.74 $\pm$ 0.11 & 22.25 $\pm$ 0.08 & 22.29 $\pm$ 0.06  & 2.6 & -0.02 & 2 \\ 
FDSLSB10 & 55.7587 & -35.3283 & 0.48 $\pm$ 0.14 & -1.0 $\pm$ 7.3 & 20.7 $\pm$ 0.4 & 7.3 $\pm$ 2.6 & 0.4 $\pm$ 0.4 & 23.79 $\pm$ 1.46 & 21.90 $\pm$ 0.28 & 21.49 $\pm$ 0.20 & 21.04 $\pm$ 0.12  & 2.0 & -0.01 & 2 \\ 
FDSLSB11 & 55.8614 & -35.4978 & 0.74 $\pm$ 0.06 & 18.5 $\pm$ 3.3 & 21.8 $\pm$ 0.2 & 1.8 $\pm$ 0.4 & 0.7 $\pm$ 0.4 & 24.19 $\pm$ 0.21 & 23.48 $\pm$ 0.15 & 22.90 $\pm$ 0.10 & 22.22 $\pm$ 0.05  & 2.7 & -0.00 & 1 \\ 
FDSLSB12 & 55.2824 & -35.4126 & 0.71 $\pm$ 0.04 & 8.7 $\pm$ 2.6 & 20.6 $\pm$ 0.2 & 2.3 $\pm$ 0.4 & 0.9 $\pm$ 0.3 & 22.91 $\pm$ 0.11 & 21.83 $\pm$ 0.07 & 21.59 $\pm$ 0.06 & 21.38 $\pm$ 0.05  & 2.8 & -0.00 & 1 \\ 
FDSLSB13 & 55.3112 & -35.4510 & 0.76 $\pm$ 0.05 & -51.1 $\pm$ 3.3 & 19.9 $\pm$ 0.2 & 4.1 $\pm$ 0.9 & 0.8 $\pm$ 0.4 & 22.10 $\pm$ 0.17 & 20.97 $\pm$ 0.08 & 20.58 $\pm$ 0.06 & 20.29 $\pm$ 0.05  & 2.4 & 0.00 & 4 \\ 
FDSLSB14 & 56.1882 & -35.2697 & 0.76 $\pm$ 0.13 & -20.2 $\pm$ 6.8 & 20.4 $\pm$ 0.4 & 7.6 $\pm$ 2.6 & 0.3 $\pm$ 0.4 & 22.89 $\pm$ 1.09 & 21.80 $\pm$ 0.42 & 21.14 $\pm$ 0.23 & 21.08 $\pm$ 0.21  & 1.6 & -0.02 & 1 \\ 
FDSLSB15 & 55.9851 & -35.8301 & 0.71 $\pm$ 0.08 & -25.3 $\pm$ 4.7 & 21.0 $\pm$ 0.3 & 3.7 $\pm$ 1.0 & 1.0 $\pm$ 0.4 & 24.15 $\pm$ 0.77 & 22.50 $\pm$ 0.20 & 21.95 $\pm$ 0.13 & 22.08 $\pm$ 0.12  & 2.7 & -0.01 & 1 \\ 
FDSLSB16 & 56.2082 & -35.5355 & 0.96 $\pm$ 0.06 & -39.0 $\pm$ 3.6 & 21.2 $\pm$ 0.2 & 2.5 $\pm$ 0.6 & 0.4 $\pm$ 0.4 & 24.86 $\pm$ 0.92 & 22.67 $\pm$ 0.16 & 22.16 $\pm$ 0.11 & 22.15 $\pm$ 0.09  & 2.2 & 0.01 & 1 \\ 
FDSLSB17 & 56.1431 & -35.1818 & 0.78 $\pm$ 0.02 & -38.2 $\pm$ 1.4 & 15.4 $\pm$ 0.1 & 11.9 $\pm$ 1.5 & 1.7 $\pm$ 0.3 & 18.35 $\pm$ 0.06 & 16.82 $\pm$ 0.03 & 16.20 $\pm$ 0.03 & 15.90 $\pm$ 0.04  & 3.3 & 0.01 & 1 \\ 
FDSLSB18 & 55.8573 & -35.2292 & 0.35 $\pm$ 0.13 & -1.6 $\pm$ 6.9 & 21.2 $\pm$ 0.4 & 5.4 $\pm$ 1.8 & 0.6 $\pm$ 0.4 & 24.13 $\pm$ 0.79 & 22.62 $\pm$ 0.23 & 22.03 $\pm$ 0.14 & 21.46 $\pm$ 0.08  & 2.4 & 0.03 & 1 \\ 
FDSLSB19 & 49.9763 & -36.4139 & 0.92 $\pm$ 0.03 & 39.4 $\pm$ 2.1 & 20.1 $\pm$ 0.2 & 2.2 $\pm$ 0.4 & 0.9 $\pm$ 0.3 & -1.00 $\pm$ -1.00 & 21.39 $\pm$ 0.06 & 21.00 $\pm$ 0.05 & 20.83 $\pm$ 0.04  & 2.9 & 0.02 & 1 \\ 
FDSLSB20 & 49.8691 & -36.8686 & 0.82 $\pm$ 0.04 & -32.3 $\pm$ 2.8 & 21.0 $\pm$ 0.2 & 2.0 $\pm$ 0.4 & 1.0 $\pm$ 0.3 & -1.00 $\pm$ -1.00 & 22.51 $\pm$ 0.10 & 22.01 $\pm$ 0.07 & 21.70 $\pm$ 0.05  & 2.7 & 0.01 & 1 \\

... & & & & & & & & & & & & & \\

\end{tabular}}
\end{table*}
\end{landscape}
\twocolumn
\end{appendix}
\end{document}